\documentclass[twocolumn]{aa}
\usepackage{graphicx}
\usepackage{txfonts}
\begin{document} 
\newcommand{\magcir}{\ \raise -2.truept\hbox{\rlap{\hbox{$\sim$}}\raise5.truept 
 	\hbox{$>$}\ }}	 
\newcommand{\mincir}{\ \raise -2.truept\hbox{\rlap{\hbox{$\sim$}}\raise5.truept 
 \hbox{$<$}\ }}  

  \title{The Great Observatories Origins Deep Survey} 
  \subtitle{VLT/FORS2 Spectroscopy in the GOODS-South Field: Part II}   
 
  \author{E. Vanzella\inst{1} 
     \and 
      S. Cristiani\inst{1} 
     \and  
      M. Dickinson\inst{2} 
     \and  
      H. Kuntschner\inst{3}      
     \and  
      M. Nonino\inst{1}
     \and  
      A. Rettura\inst{5,6}   
     \and
      P. Rosati\inst{5} 
     \and 
      J. Vernet\inst{5}  
     \and
     \\
      C. Cesarsky\inst{5}    
     \and 
      H. C. Ferguson\inst{4} 
     \and
      R.A.E. Fosbury\inst{3} 
     \and 
      M. Giavalisco\inst{4}   
     \and
      A. Grazian\inst{9}
     \and
      J. Haase\inst{3}
     \and
      L. A. Moustakas\inst{7} 
     \and
     \\
      P. Popesso\inst{5}
     \and
      A. Renzini\inst{8}  
     \and  
      D. Stern\inst{7} 
     \and
     the GOODS Team 
     } 
 
  \institute{
       INAF - Osservatorio Astronomico di Trieste, Via G.B. Tiepolo 11,
       40131 Trieste, Italy.
     \and
       National Optical Astronomy Obs., P.O. Box 26732, Tucson, AZ 85726.
     \and
       ST-ECF, Karl-Schwarzschild Str. 2, 85748 Garching, Germany.
     \and
       Space Telescope Science Institute, 3700 San Martin Drive,
       Baltimore, MD 21218.
     \and
       European Southern Observatory, Karl-Schwarzschild-Strasse 2,
Garching, D-85748, Germany.
     \and
       Universite' Paris-Sud 11, Rue Georges Clemenceau 15, Orsay, F-91405, France
     \and
       Jet Propulsion Laboratory, California Institute of Technology,
       MS 169-506, 4800 Oak Grove Drive, Pasadena, CA 91109
     \and
       INAF - Astronomical Observatory of Padova, Vicolo dell'Osservatorio 5, 
       I - 35122 Padova -ITALY 
     \and
       INAF - Osservatorio Astronomico di Roma, Via Frascati 33, 
       I-00040 Monteporzio Roma, Italy  
     \thanks{Based on observations made at the European Southern
Observatory, Paranal, Chile (ESO programme 170.A-0788 {\it The Great
Observatories Origins Deep Survey: ESO Public Observations of the
SIRTF Legacy/HST Treasury/Chandra Deep Field South.}) }
       } 
 
  \offprints{E. Vanzella, \email{vanzella@oats.inaf.it}}

  \date{Received ; accepted }
 
 \abstract
 {}
 {We present the second campaign of the ESO/GOODS program of spectroscopy
 of faint galaxies in the GOODS-South field.}
 {Objects were selected as candidates for VLT/FORS2 observations 
  primarily based on the expectation that the detection and measurement 
  of their spectral features would benefit from the high throughput and 
  spectral resolution of FORS2. The reliability of the redshift estimates 
  is assessed using 
  the redshift-magnitude and color-redshift diagrams and comparing the results with public data.}
 {807 spectra  
 of 652 individual targets have been obtained in service mode with 
 the FORS2 
 spectrograph at the ESO/VLT, providing 501 redshift 
 determinations.
 The typical redshift uncertainty is estimated to be 
 $\sigma_z \simeq 0.001$. 
 Galaxies have been 
 selected adopting three different color criteria and using the photometric redshifts.
 The resulting redshift distribution typically spans two redshift domains:
 from z=0.5 to 2 and z=3.5 to 6.2. In particular, 94
$B_{435}$-,$V_{606}$-,$i_{775}$-"dropout" 
 Lyman break galaxies have been observed, yielding redshifts for 65 objects in the
 interval 3.4$<$z$<$6.2. Three sources have been serendipitously discovered in the
 redshift interval 4.8$<$z$<$5.5.
 Together with the previous release, 930 sources have now been observed 
 and 724 redshift determinations have been carried out.
 The reduced spectra and the derived redshifts are released to the community  
 through the ESO web page $\it{http://www.eso.org/science/goods/}$
 \thanks{The catalog (Table~\ref{tab:tblspec}) is available in electronic form
 at the CDS via anonymous ftp to cdsarc.u-strasbg.fr (130.79.128.5)
 or via http://cdsweb.u-strasbg.fr/cgi-bin/qcat?J/A+A/}.
 Large scale structures are clearly detected at $z \simeq 0.666, 0.734, 1.096, 1.221, 1.300$,  
 and $1.614$. A sample of 34 sources with tilted [O\,{\sc ii}]3727 emission
 has been identified, 32 of them in the redshift range 0.9$<$z$<$1.5. 
}
 {}

 \maketitle 
% 
%________________________________________________________________ 
 
\section{Introduction} 
The Great Observatories Origins Deep Survey (GOODS) is a public, 
multi-facility project that aims to answer some of the most profound 
questions in cosmology: how did galaxies form and assemble their 
stellar mass? When was the morphological differentiation of galaxies 
established and how did the Hubble Sequence form? How did Active
Galactic Nuclei (AGN) form and 
evolve, and what role do they play in galaxy evolution? How much do 
galaxies and AGN contribute to the extragalactic background light? Is 
the expansion of the universe dominated by a cosmological constant? A 
project of this scope requires large and coordinated efforts from many 
facilities, pushed to their limits, to collect a database of 
sufficient quality and size for the task at hand. It also requires 
that the data be readily available to the worldwide community for 
independent analysis, verification, and follow-up. 
 
The program targets 
two carefully selected fields, the Hubble Deep Field North (HDF-N) and 
the Chandra Deep Field South (CDF-S), with three NASA Great 
Observatories (HST, Spitzer and Chandra), ESA's XMM-Newton, and a wide 
variety of ground-based facilities. The area common to 
all the observing programs is 320 arcmin$^2$, equally divided between 
the North and South fields. For an overview of GOODS, see \cite{dick03},  
\cite{renz03} and \cite{giava04a}.  
In the last five years
the CDF-S has been the target of several spectroscopic campaigns
(\cite{crist00}, \cite{croom01}, \cite{bunk03}, \cite{stan04},
\cite{stro04}, \cite{vanderwel04}, \cite{dick04}, \cite{szo04},
\cite{fevre05}, \cite{vanz05}). 

This is the second paper in a series presenting the results of the GOODS
spectroscopic program carried out with the VLT/FORS2 spectrograph.
For a full description of its aims we refer to the first
paper (\cite{vanz05}, RUN1 hereafter).
 
Here we recall that the ESO/GOODS spectroscopic program is designed to observe 
all galaxies for which VLT optical spectroscopy is likely to allow the redshift 
determination. 
The program makes full use of the VLT instrument capabilities 
(FORS2 and VIMOS), 
matching targets to instrument and disperser combinations in order to 
maximize the effectiveness of the observations. The magnitude limits 
and selection bandpasses to some extent depend on the instrumental 
setup being used. The aim is to reach mag~$\sim24-25$ with adequate S/N, with 
this limiting magnitude being in the B band for objects observed with 
the VIMOS LR-Blue grism, in the V band for those observed in the VIMOS 
LR-Red grism, and in the z band for the objects observed 
with FORS2.  
 
The second FORS2 spectroscopic campaign (17 masks, RUN2 hereafter) in the 
Chandra Deep Field South, was carried out 
in the period fall 2003 - early 2004 in service mode. 
New FORS2 observations were performed in December 2004 (6 masks, RUN3) mainly
focused on color-selected Lyman break ``dropout'' targets and 5 more masks will be 
completed before February 2006 (RUN4) mainly dedicated to sources detected 
at 24$\mu$m with the {\it Spitzer Space Telescope} MIPS instrument.
These data will be described in a forthcoming paper.
The VIMOS spectroscopic survey in the GOODS-S field is started and will produce
hundreds of redshift determinations, mainly in the redshift range 0$<$z$\leq$3.5.

The paper is organized as follows. Sect. 2 describes the target selection,
while Sect. 3 describes the observations and data reductions.
The redshift determination is presented in Sect. 4. In Sect. 5 we discuss
the data and in Sect. 6 the conclusions are presented.
Throughout this paper the magnitudes are given in the AB system (\cite{oke77})
(AB~$\equiv 31.4 - 2.5\log\langle f_\nu / \mathrm{nJy} \rangle$),
and the ACS F435W, F606W, F775W, and F850LP filters are designated
hereafter as $B_{435}$, $V_{606}$, $i_{775}$ and $z_{850}$, respectively.
We assume a cosmology with $\Omega_{\rm tot}, \Omega_M, \Omega_\Lambda = 1.0, 0.3, 0.7$
and $H_0 = 70$~km~s$^{-1}$~Mpc$^{-1}$.
 
%__________________________________________________________________ 
 
\section{Target Selection} 

\begin{figure} 
 \centering
 \includegraphics[width=9cm,height=9.5cm]{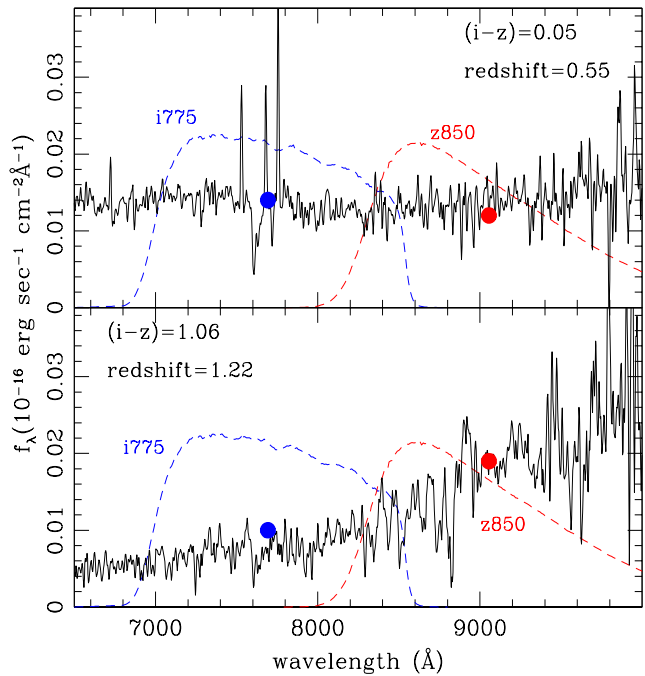} 
 \caption{A comparison between two 1-D calibrated spectra and the ACS photometry in 
the $i_{775}$ and $z_{850}$ bands (filled circles, the shapes of the transmission of the filters 
are shown).
In the top panel one of the bluest galaxies present in the FORS2 spectroscopic sample 
with ($i-z$)=0.05 is shown, while in the bottom panel a red elliptical galaxy is shown. 
The photometric data points have been calculated at the central wavelengths of the filters 
(7693\AA~ and 9055\AA~ for the $i_{775}$ and $z_{850}$ bands, respectively). 
The uncertainty of the photometric data is smaller than the size of the filled circles.}
\label{fig:test_flux}
\end{figure}

Galaxies were selected as candidates for FORS2 observations 
primarily based on the expectation that the detection and measurement 
of their spectral features would benefit from the high throughput,
moderately-high spectral resolution, and reduced long-wavelength fringing
of FORS2 relative to other instrument options such as VIMOS.
In particular, the main spectral emission and absorption features for 
galaxies at $0.8 < z < 2.0$ appear at very red optical wavelengths 
($7000\AA < \lambda < 1\mu$m).  Similarly, very faint Lyman break 
galaxies at $z \gtrsim 4$, selected as $B_{435}$, $V_{606}$ and 
$i_{775}$--dropouts from the GOODS ACS photometry, also benefit greatly 
from the red throughput and higher spectral resolution of FORS2.
 
In practice, several categories of object selection criteria were used to 
ensure a sufficiently high density of targets to 
efficiently populate masks. These criteria were: 
 
\begin{enumerate} 
 
\item{Primary catalog:  $(i_{775}-z_{850}) > 0.6$ and $z_{850} < 25$. 
This should ensure redshifts $z \gtrsim 0.7$ for ordinary early-type galaxies
(whose strongest features are expected to be absorption lines),
and higher redshifts for intrinsically bluer galaxies likely to have emission 
lines.}
 
\item{Secondary catalog:  $0.45 < (i_{775}-z_{850}) < 0.6$ and $z_{850} < 25$.}
\item{Photometric-redshift sample:  $1<$$z_{\rm phot}$$<2$ and $z_{850} < 25$, 
from \cite{moba04}.} 
 
\item{$B_{435}$, $V_{606}$ and $i_{775}$--dropouts color selected Lyman break 
galaxy candidates (see \cite{giava04b} and \cite{dick04}).}
 
\item{A few miscellaneous objects, including host galaxies of
supernovae detected in the GOODS ACS observing campaign.}
 
\end{enumerate} 
 
The targets were selected from a preliminary catalog based on the
v1.0 public release of the GOODS ACS images. This version includes
all five epochs of the GOODS ACS data 
\footnote{{\it ftp://archive.stsci.edu/pub/hlsp/goods/catalog$\_$r1/}}, 
and is a significant improvement
on the previous, 3-epoch v0.5 release that was used to select targets
for the first FORS2 observations (RUN1, \cite{vanz05}).
For this paper and data release, the objects observed with FORS2
have been matched to the public release ACS catalog version r1.1z,
also based on the 5-epoch v1.0 ACS images. The r1.1z catalog
is based on the r1.0z SExtractor run, and merely corrects errors and
omissions in the r1.0z catalog files.

When designing the masks, we generally 
tried to avoid observing targets that had already been observed in other 
redshift surveys of this field, namely the K20 survey of \cite{cimatti02}  
and the survey of X-ray sources by \cite{szo04}. 

807 spectra of 652 individual targets have been extracted 
from the RUN2 (multiple observations have been performed,
especially for the high redshift candidates).
Out of these 652 targets, 178 are from the primary catalog, 117 are from
the secondary catalog, 141 are from the photometric redshift selection,
94 are from the Lyman break sample, and 3 are from miscellaneous list. 
The remaining 119 sources have been serendipitously identified, due to:
a) sources randomly in the slit other than the target or 
b) sources put in the slit in the situation where no targets were available
or c) relatively bright objects put in the slit for the alignment of the mask.

\begin{figure} 
 \centering 
 \includegraphics[width=9cm,height=9cm]{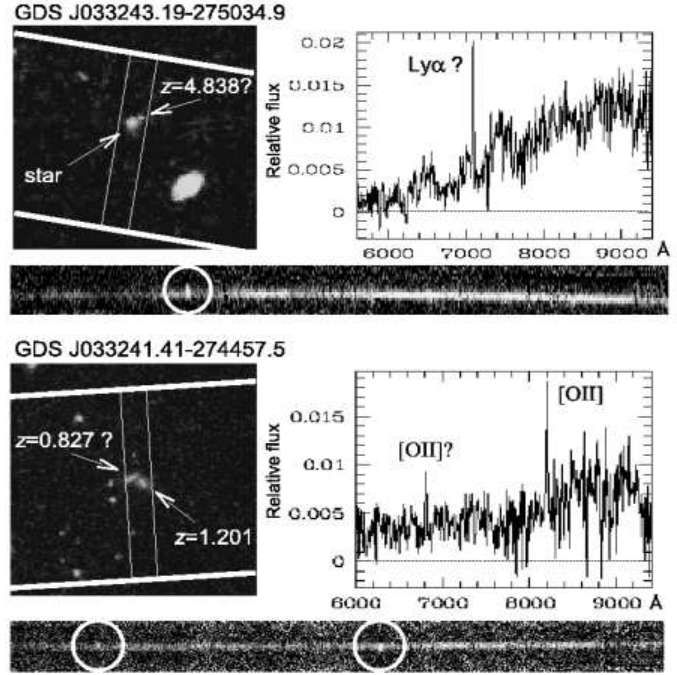}
 \caption{Two examples of objects at different redshift superimposed in the slit are shown.
In the top panel the spectrum of a star with an emission line is shown, in the
two dimensional spectrum the line (marked with a circle) is detected also $\sim$0.5'' 
beyond the trace of the star. In the color image the presence of
the second source is visible in the $i_{775}$ and $z_{850}$ bands and disappears in the
$B_{435}$ and $V_{606}$ bands. We have assigned tentatively $z$=4.838 (QF=''C''). 
In the lower panel a similar case of two close sources is shown. The two emission 
lines in the spectrum are positioned in the top and bottom part of the main trace, 
consistent with the geometry of the system shown in the ACS color image. 
In both cases the serendipitous source is not present in the ACS catalog (v1.1).} 
\label{fig:MIXED} 
\end{figure} 

The total number of individual sources observed in 
the RUN1 + RUN2 is 930 (1203 spectra reduced) with 724 redshift determinations.
The spectroscopic database presented here is incomplete:
none of the above listed categories has been exhaustively observed,
nor any GOODS subarea has been fully covered.

\section{Observations and Data Reduction}
\begin{table}
\centering \caption{Journal of the FORS2 observations (RUN2).}
\begin{tabular}{lccc}
\hline \hline
 Mask ID & Date & exp.time (s)\\
\hline
 914250 & Aug. 2003         & 17$\times$1200 \cr
 905513 & Sept. 2003        & 18$\times$1200   \cr
 943018 & Sept. 2003        & 12$\times$1200   \cr
 924345 & Sept. 2003        & 12$\times$1200   \cr
 945143 & Sept. - Oct. 2003 & 12$\times$1200 + 3$\times$1000  \cr
 992438 & Oct. - Dec. 2003  & 12$\times$1200 \cr
 985931 & Nov. 2003         & 12$\times$1200 + 2$\times$120 \cr 
 990204 & Dec. 2003         & 12$\times$1200  \cr 
 904509 & Dec. 2003         & 12$\times$1200  \cr 
 991435 & Dec. 2003         & 12$\times$1200   \cr  
 935030 & Dec. 2003         & 12$\times$1200   \cr 
 951937 & Dec. 2003         & 12$\times$1200 + 1100 + 500   \cr 
 960930 & Dec. 2003         & 12$\times$1200   \cr 
 961839 & Jan. 2004         & 12$\times$1200  \cr 
 932802 & Jan. 2004         & 12$\times$1200   \cr 
 993304 & Jan. 2004         & 12$\times$1200   \cr 
 951526 & Feb. 2004         & 3$\times$1200 \cr 
\hline 
\label{tab:tblobs} 
\end{tabular} 
\end{table} 

The VLT/FORS2 spectroscopic observations were carried out  
in service mode during several nights at the end of 2003 and the beginning 
of 2004.  
A summary is presented in Table~\ref{tab:tblobs}. 
In all cases the $300I$ grism was used as dispersing element without  
order-separating filter. 
This grism provides a scale of roughly 3.2\AA~pix$^{-1}$. The nominal 
resolution of the configuration was 
R=$\lambda/\Delta\lambda$=660, which corresponds to
13{\AA} at 8600{\AA}. The spatial scale of FORS2 is $0.126\arcsec$/pixel.
The slit width was always $1\arcsec$. 
Dithering of the targets along the slits was applied typically with
steps of 0,$\pm$8 pixels,
in order to effectively improve the sky and fringe subtraction, and remove 
CCD blemishes. 

\subsection{\it Data Reduction} 

Data were reduced with a semi-automatic pipeline  
that we have developed on the basis of the MIDAS package (\cite{eso_midas}), 
using commands of the LONG and MOS contexts. 
The main procedures have been described in the previous paper (\cite{vanz05}).

In the cases of multiple observations of the same source
in different masks, the one dimensional spectra have been co-added
weighing according to the exposure time, the seeing condition and the
resulting quality of each extraction process (defects present in the CCD, object too
close to the border of the slit, etc.). A visual check of the two
dimensional frames has been performed (in some cases the two
dimensional spectra have also been co-added, in order to improve and guide the
visual inspection).

We emphasize here that we opted to observe the science targets {\em without} an 
order-sorting filter, implying deleterious effects to the flux calibration.
The second order overlap becomes important at wavelengths above
$\sim$8000\AA~depending on the color of the target.
In Figure~\ref{fig:test_flux} the comparison between two 1-D calibrated spectra
(one blue ($i-z$)$\sim$0 and one red ($i-z$)$\sim$1)
and the correspondent ACS photometry is shown.
The photometric values in the $i$ and $z$ bands are marked with two filled circles 
and are consistent with the derived spectral behavior.
For the red objects that dominate the FORS2 target selection, we felt that the
improved wavelength coverage more than compensates for the partial
unreliability of the flux calibration. Due to both this second order
light and uncertain slit losses, we caution against using the
calibrated fluxes for scientific purposes.
Fluxes in the released 1-D spectra are given in units of $10^{-16}$ erg s$^{-1}$
cm$^{-2}$ \AA$^{-1}$.

\begin{figure} 
 \centering 
 \includegraphics[width=8.7cm,height=9.5cm]{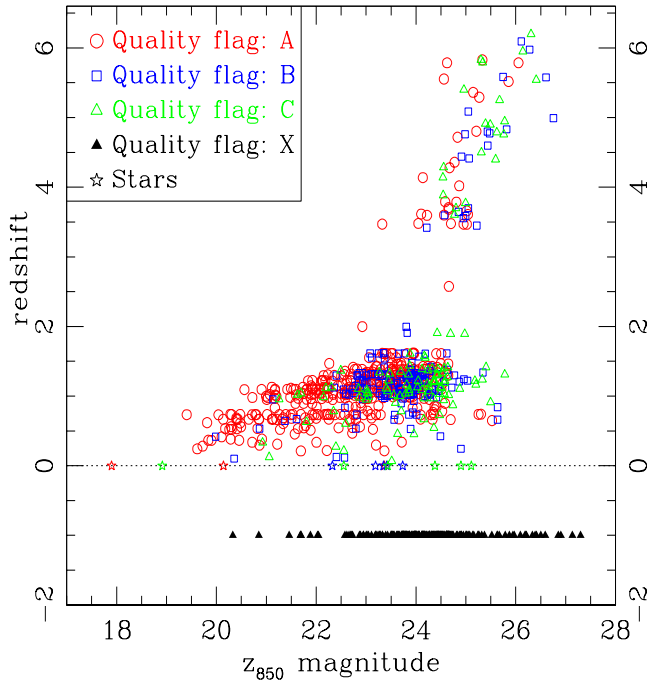}
 \caption{Spectroscopic redshift versus magnitude for the FORS2 catalog  
(quality flag ``A'', ``B'', ``C'' and ``X''). Stars are 
denoted by star-like symbols at zero redshift. Inconclusive spectra 
are placed at $z=-1$. The gap in the redshift interval 2$<$z$<$3.5 is due
to the spectral coverage adopted ($\sim$ 5800\AA-10000\AA) and will be (partly) 
filled with the VIMOS spectroscopic observations.}  
\label{fig:z_vs_mag} 
\end{figure} 
\begin{figure}
 \centering
 \includegraphics[width=8.7cm,height=9.5cm]{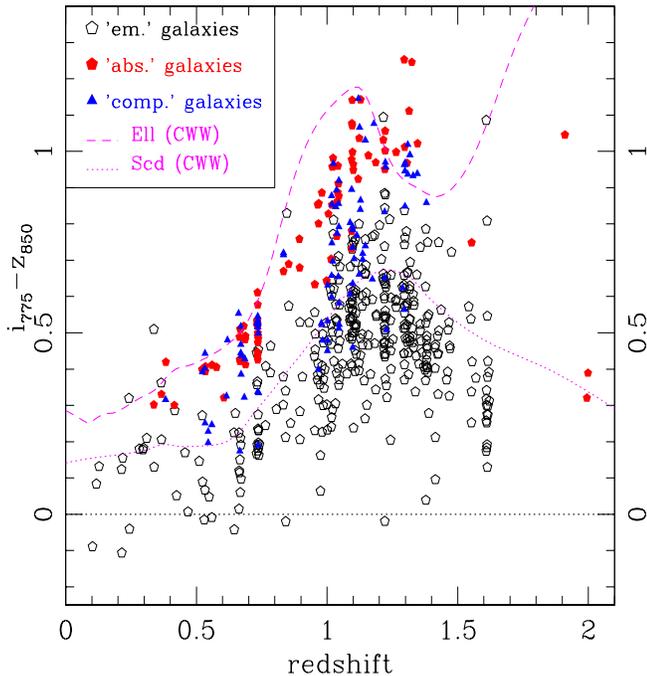} 
 \caption{Color-redshift diagram of the spectroscopic sample,
presenting only galaxies with quality flag ``A'' and ``B''.
Filled pentagons symbols are objects identified with absorption
features only (``abs.'' sources),  
while open pentagons are objects showing only  
emission lines (``em.'' sources).  
The intermediate cases are shown by filled triangles (``comp.'' sources). 
The long-dashed line and the short dashed line show the colors of a non-evolving  
$L^{\star}$ elliptical galaxy and an Scd galaxy, respectively, using the spectral
templates of CWW \cite{cole80}.} 
\label{fig:i_zVSzspec} 
\end{figure} 

\section{Redshift Determination}
Spectra of 652 individual objects have been extracted from RUN2. 
From them we have determined 501 redshifts.  
In the large majority of the cases the redshift has been determined through the 
identification of prominent features of galaxy spectra: 
depending on the redshift and the nature of the source the 4000\AA\ break, Ca H and K, 
g-band, MgII 2798-2802, AlII 3584,
Ly$\alpha$, Si\,{\sc ii} 1260.4\AA,  O\,{\sc i} 1302.2\AA, C\,{\sc ii} 1335.1\AA, 
Si\,{\sc iv} 1393.8,1402.8\AA, Si\,{\sc ii} 1526.7\AA, C\,{\sc iv} 1548.2, 1550.8\AA~in 
absorption and Ly$\alpha$, [O\,{\sc ii}]3727, [O\,{\sc iii}]5007, H$\beta$, H$\alpha$
in emission. 
The redshift estimation has been performed cross-correlating the observed 
spectrum with templates of different spectral types (S0, Sa, Sb, Sc, Elliptical, 
Lyman Break, etc.), using the $rvsao$ package in the IRAF environment.
The redshift identifications are summarized in Table~\ref{tab:tblspec} and 
are available at the URL $\it{http://www.eso.org/science/goods/}$. 
 
In Table~\ref{tab:tblspec},  
the column {\em ID} contains the target identifier, that is constructed out of the  
target position (e.g., $GDS~J$033206.44-274728.8) where GDS stands 
for {\bf G}OO{\bf D}S {\bf S}outh.  
The coordinates are based on the GOODS v1.1 astrometry.
The v1.1 release is based upon the v1.0 SExtractor run, and merely corrects 
errors and omissions in the v1.0 catalog files.  
The cataloged sources are identical, in both number and ordering, to the v1.0 
release. 
The columns $z_{850}$ and ($i_{775}$-$z_{850}$) list the magnitude 
(SExtractor ``MAG$\_$AUTO'') and the color (SExtractor ``MAG$\_$ISO'') of the sources
derived from the catalog v1.1. The color has been measured through isophotal 
apertures defined in the $z_{850}$ band image (as done in \cite{dick04} and
\cite{giava04b}).

The {\em quality} flag (QF hereafter), indicates the reliability of the
redshift determination. As described in the previous work (\cite{vanz05}, RUN1),
the QF has been divided in three categories: ``A'', ``B'' and ``C''.
An estimation of the confidence level associated to each class ``A'', ``B'' and ``C''
can be derived analyzing the FORS2 measurements in common with independent spectroscopic
estimations available in literature. 
This has been done in the previous paper (RUN1) where 39 sources have been analyzed
and in Sect. 5.1 of the present work (98 more sources, see below). 
In this way the sample of FORS2 measurements in common with independent spectroscopic
surveys counts 137 galaxies, in which we find 0, 1 and 4 FORS2 wrong redshifts for classes ``A'', ``B'' 
and ``C'', respectively.
In this way at 1$\sigma$ (\cite{gehrels86}) the confidence level of the ``A'', ``B'' and ``C''
categories turns out to be $\simeq$ 98$\%$, $\simeq$ 97$\%$ and 93$\%$.

There are 291 objects classified with quality
``A'', 119 with quality ``B'', 91 with ``C'', and 151 with ``X'', an inconclusive spectrum.
 
The flag "{\em class}" groups the objects for which emission line(s) (``em.''), 
absorption-line(s) (``abs.'') or both (``comp.'') are detected in the spectrum. 
In the present catalog, three sources have been classified as stars. 
 
In 30$\%$ of the cases the redshift is based only on one emission line, usually 
identified as [O\,{\sc ii}]3727 or Ly$\alpha$.  
In these cases the continuum shape, the presence of breaks, the 
absence of other spectral features in the observed spectral range and the broad band  
photometry are particularly important in the evaluation. The quality for these
sources ranges from ``A'' to ``C'' depending on the additional information 
described above (35$\%$ of the sample with a single emission line have
QF=''A'', with a mean redshift $<z>$=1.21$\pm$0.2).  
 
The {\em comments} column contains additional information relevant 
to the particular observation. The most common ones summarize  
the identification of the principal lines, the inclination of an emission  
line due to internal kinematics, the weakness of the signal  
(``faint''), the low S/N of the extracted spectrum (``noisy''),  the apparent absence
of spectroscopic lines (``featureless continuum''), etc.
 
In few cases the spectrum extracted is the combination of more than
one source in the slit and where possible the redshifts of the
``components'' have been estimated.  In the RUN1 + RUN2 spectroscopic
data, 11 sources in the GOODS-S field are not present in the ACS
photometric catalog v1.1. Six of them have a redshift estimation (an
example is shown in Figure~\ref{fig:MIXED}).  Three out of six appear
to be emission line objects whose continuum is too faint and has not
been detected in the ACS catalogs. The other seven sources are outside
the ACS area.

\begin{table*}
\centering \caption{Spectroscopic redshift catalog. $\dag$}
\begin{tabular}{lcccccl}
\hline \hline 
  ID(v1.0)       & $z_{850}$      & $(i_{775}-z_{850})$  & zspec         &class. & Quality & comments \\ 
\hline 
  GDS~J033245.99-275108.3  &     23.48  &0.47   &1.238 &em.    &  B    &[O\,{\sc ii}]3727  \cr 
  GDS~J033246.04-274929.7  &     26.06  &1.77   &5.787 &em.    &  A    &LyA (faint continuum) \cr 
  GDS~J033246.05-275444.8  &     21.49  &0.53   &0.733 &abs.   &  A    &CaH,g-band,H$\beta$,Mg,CaFe \cr 
  GDS~J033246.16-274752.3  &     24.46  &0.43   &1.221 &em.    &  B    &[O\,{\sc ii}]3727  \cr 
\hline   
\multicolumn{6}{l}
{$\dag$ This table is available in its entirety via $\it{http://www.eso.org/science/goods/}$.}\\
\multicolumn{6}{l}
{A portion is shown here for guidance regarding its form and content.}\\
\label{tab:tblspec} 
\end{tabular} 
\end{table*} 

\section{Discussion}

In the following, if not specified, we consider the entire FORS2 sample, including
both RUN1 and RUN2. This sample is summarized in Table~\ref{tab:matrix} where the sources
are divided into different selection categories (see Sec. 2) and by redshift
(or whether a redshift could be determined).
The distribution of the quality flags is also tabulated.
\begin{figure} 
 \centering 
 \includegraphics[width=9cm,height=9cm]{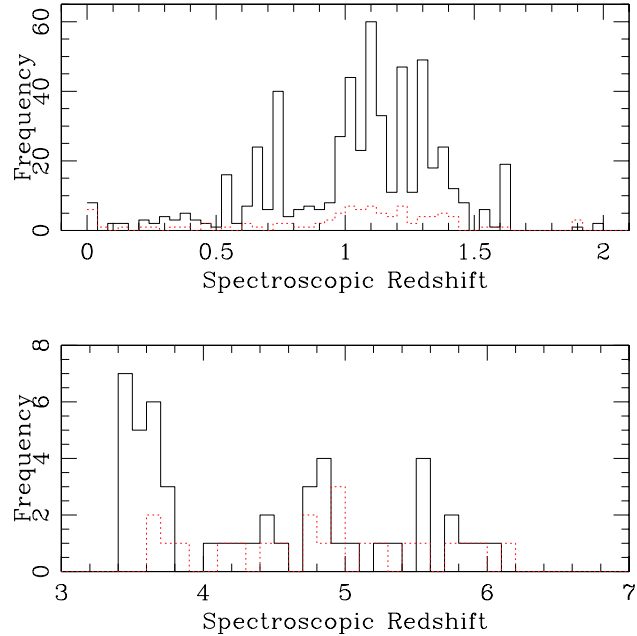}
 \caption{Redshift distribution according to the selection functions described in Sect. 2
 for the spectroscopic sample with  
 quality A, B (solid line) and C (dotted line). In the top and bottom panels 
 the sources at z$<$2 ($dz=0.04$) and z$>$2 ($dz=0.1$) are shown, respectively.}
\label{fig:zdistr} 
\end{figure} 

\subsection{Reliability of the redshift - comparison with public data}

\begin{figure*} 
 \centering 
 \includegraphics[width=13cm,height=11cm]{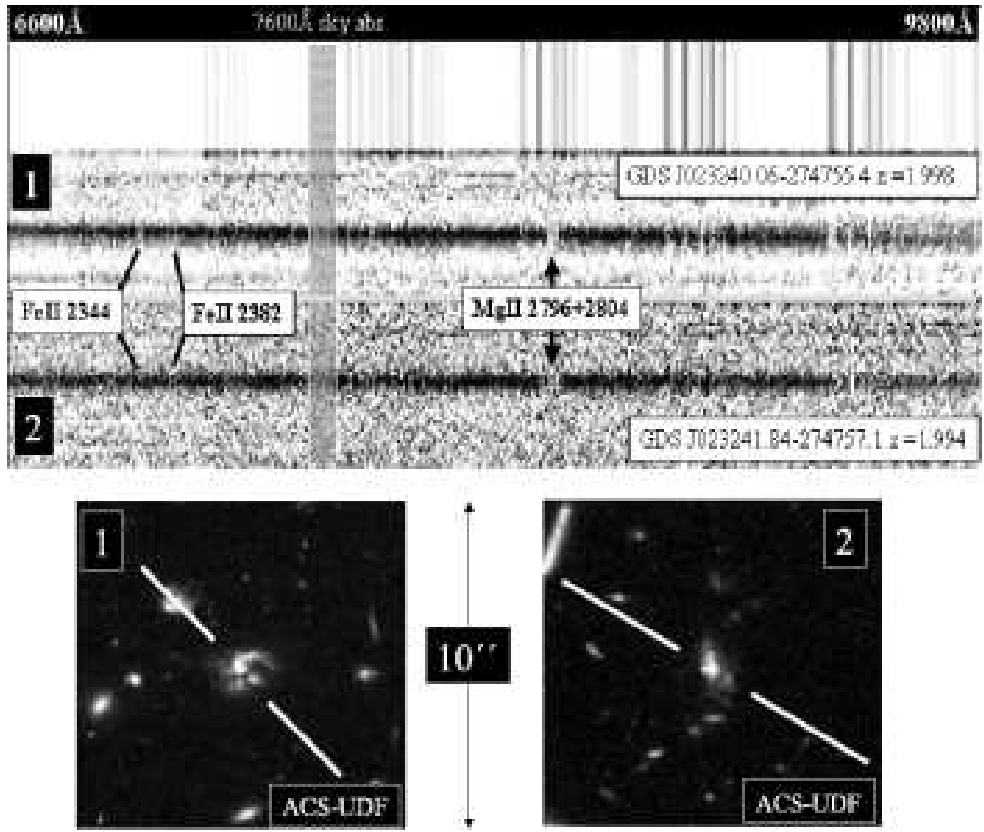}
 \caption{Two galaxies at z$\simeq$2 identified with absorption lines Mg\,{\sc ii} 2798,2802\AA~and 
 [Fe\,{\sc ii}] 2344,2383\AA~ (see text). In the top part the two dimensional spectra of the sources
 and the sky lines are shown. The bottom part shows the ACS-UDF color images. The solid lines
 indicate the orientation of the slit.} 
 \label{fig:GAL_Z2} 
\end{figure*} 

A practical way to assess the reliability of the redshifts reported in
Table~\ref{tab:tblspec} is to compare the present results with
independent measurements from other surveys. In the last five years
the CDF-S has been the target of several spectroscopic campaigns
(the surveys with the number of redshifts in parenthesis used in the 
comparison are here reported:
\cite{crist00} (5), \cite{croom01} (29), \cite{bunk03} (1),
\cite{stan04} (3), \cite{stro04} (14), \cite{vanderwel04} (6),
\cite{dick04} (1), \cite{szo04} (124),
\cite{fevre05} (748), \cite{vanz05} (234)).
Making use of a publicly available
master compilation of all spectroscopic redshifts in the GOODS/CDF-S
region (Rettura et al. in preparation, available at the URL {\it
http://www.eso.org/science/goods/spectroscopy/CDFS$\_$Mastercat/} we
have been able to compare our redshift determinations with the
existing data in the literature.

There are 98 objects in common with the present second release of the
FORS2 GOODS survey (RUN2).
For 87 cases out of 98 (89$\%$) the agreement is very good, with a mean difference
$<z_{FORS2_{RUN2}} - z_{CDF-S}>$= 0.0042 $\pm$ 0.0095.

15 objects have a redshift determination both in RUN1 and RUN2. The
distribution of the redshift differences has a median
$|z_{FORS2_{RUN2}} - z_{FORS2-{RUN1}} | = 0.0002$ and a difference
between the 82 and 18 percentile of $2.6 \cdot 10^{-3}$.
Assuming equipartition of the redshift uncertainties between RUN1 and
RUN2 we derive a typical error on the redshift determinations
in the FORS2 GOODS spectroscopy of $\sigma_z \simeq 0.001$.

Ten cases show ``catastrophic'' discrepancies between the RUN2 and  
the K20, \cite{szo04} and the VVDS surveys, 
i.e. $|z_{FORS2_{RUN2}} - z_{CDF-S}|$ greater
than 0.08.  In order to compare the redshift estimations we recall
here which is the quality level adopted by other authors. In the K20
survey the QF adopted is 1, 0 or -1 if the redshift determination is
solid, tentative or unconclusive, respectively.
In \cite{szo04} the QF=3 indicates reliable redshift determination
with unambiguous X-ray counterpart, QF=2 corresponds to a reliable
redshift determination and a value of 0.0 indicates no success. QF=1
indicates the detection of $some$ feature in the spectrum (typically a
single narrow emission line). QF=0.5 is used when there is a hint of
some spectral feature.  In the VVDS, the flags 2,3,4 are the most
secure with a confidence of 75$\%$, 95$\%$ and 100$\%$
respectively. Flag 1 is an indicative measurement (50$\%$), flag 9
indicates that there is only one secure emission line, and flag 0
indicates a measurement failure with no features identified.

In the following we discuss in detail each discrepant spectrum:

\begin{enumerate}

\item{GDS~J033232.08-274155.2. This is a discrepancy with our previous
identification (RUN1) and the present one (RUN2). In the first run the
redshift determination was tentative (quality ``C'', z= 0.960) and in
the second run we derived z=1.393 (QF=''B''). However the co-addition
of the two produces a featureless continuum, we have changed the
quality to ``X''.}

\item{GDS~J033217.77-274714.9. K20 and VVDS assign redshift 0.729 and
0.731, respectively (and quality 1 and 3).  In the FORS2 spectrum
there are three objects in the slit, the GDS~J033217.77-274714.9 is a
serendipitous source at the border of the slit, its exposure time is
reduced of 50$\%$ due to the dithering process.  The continuum is
faint and a possible emission line is detected at
7522.6\AA~interpreted to be [O\,{\sc ii}]3727 at z=1.018 (QF=''C'').} 

\item{GDS~J033232.18-274534.9. K20 assigns a redshift 0.332 with quality 1. The FORS2 spectrum 
shows [O\,{\sc ii}]3727, MgI, CaHK, g-band and the Balmer Break at z=0.523 (QF=''A'').} 

\item{GDS~J033239.67-274850.6. Szokoly et al. measure redshift 3.064 with quality 3. 
Our spectrum shows a smoothed
break at $\sim$ 6000\AA~and an absorption line at 6789.0\AA, our redshift determination is
tentatively z=3.885, QF=''C''. The spectrum starts at 5600\AA, if it is at redshift 3.064, 
the most relevant spectral features are outside the spectral
coverage. 
We note that if the redshift is 3.064
the absorption line we measure at 6789.0\AA would be consistent with the Al\,{\sc ii} 1670.8\AA.} 

\item{GDS~J033240.84-275546.7. Szokoly et al. measure redshift 0.625 with quality 0.5.
Our spectrum shows a featureless continuum and starts at 5790\AA, a possible emission line is
detected at 8277.6\AA, we assign tentatively z=1.221 QF=''C''.} 

\item{GDS~J033222.44-275606.1. VVDS measure redshift 0.490 with quality 2.
The FORS2 spectrum shows a tilted emission line at 7790.3\AA~and a faint-diffuse continuum. We
assign tentatively z=1.090 (QF=''C''). We note that in the FORS2 spectrum the [O\,{\sc ii}]3727,
[O\,{\sc iii}]5007 or H$\beta$ lines at z=0.490 have not been detected.} 

\item{GDS~J033225.28-275524.2. VVDS measure redshift 0.923 with quality 1.
The FORS2 spectrum shows [O\,{\sc ii}]3727 (slightly tilted), CaHK, MgI and the Balmer Break 
at z=1.017 (QF=''A'').} 

\item{GDS~J033230.37-274008.5. VVDS measure redshift 1.083 with quality 2.
The source shows a bright continuum and [O\,{\sc ii}]3727, MgII and the NeIII lines 
at z=1.357 (QF=''A'').} 

\item{GDS~J033230.50-275312.3. VVDS measure redshift 1.427 with quality 2.
The FORS2 spectrum shows [O\,{\sc ii}]3727 (tilted), CaHK, MgII, g-band at z=1.017 QF=''A''.}

\item{GDS~J033234.82-274721.9. VVDS measure redshift 0.315 with quality 3.
In the FORS2 spectrum an emission line has been detected at 8632.3\AA, interpreted as [O\,{\sc ii}]3727 
at z=1.316 with QF=''B''.
The continuum starts at 6260\AA, and if we assume the line to be H$\alpha$ at z=0.315 
the H$\beta$ and/or [O\,{\sc iii}]5007 are not present.} 

\item{GDS~J033242.97-274649.9. VVDS measure redshift 0.831 with quality 1.
The FORS2 spectrum shows [O\,{\sc ii}]3727, CaHK, NeIII and $H\delta$ (in absorption) at z=1.036 
with QF=''A''.} 

\end{enumerate}

In summary, 7 out of 10 discrepant redshift determinations turn out to be probably correct in the FORS2 
spectroscopy, all with QF better or equal to QF=``B''. 
Of the remaining 3 sources (all with QF=''C''), one is uncertain and two are probably wrong
in the FORS2 spectroscopic identification due to the reasons described above.   

\subsection{Reliability of the redshifts - diagnostic diagrams} 

Figures~\ref{fig:z_vs_mag} and \ref{fig:i_zVSzspec} 
show the redshift-magnitude and the color-redshift distributions, 
respectively. Figure~\ref{fig:i_zVSzspec} shows the behavior
for galaxies at redshift less than 2 and quality flag ``A'' and ``B''. 
The two populations of ``emission-line'' (star-forming)  and ``absorption-line'' 
(typically elliptical) galaxies are clearly separated. 
The mean color of the absorption-line objects outline the upper envelop
of the distribution,

consistent but increasingly bluer than the colors of a non-evolving $L^{\star}$ 
elliptical galaxy (estimated integrating the spectral templates of \cite{cole80}  
through the ACS bandpasses). 
 
The emission-line objects show in general a bluer $i_{775}-z_{850}$ color and 
a broader distribution than the absorption-line sources.
The broader distribution, with some of the emission-line objects 
entering the  color region of the ellipticals,  
is possibly explained by dust obscuration, 
high metallicity or strong line emission in the $z_{850}$ band 
(for example emission lines [O\,{\sc iii}]5007, H$\beta$ at redshift 0.8, as measured
for the source GDS~J033219.53-274111.6).

\begin{table*} 
\centering \caption{Summary of the spectroscopic catalog as a function of the redshift bin
(first column), categories (from column two to six) and serendipitously identified sources (column seven).
The contribution of the different quality flags (``A'', ``B'' or ``C'') are also reported. 
A total of 930 spectra have been analyzed (RUN1 and RUN2).}
\begin{tabular}{lllllll|c} 
\hline \hline 
 z-bin  & cat. 1)$_{(A,B,C)}$ & cat. 2)$_{(A,B,C)}$  & cat. 3)$_{(A,B,C)}$ & cat. 4)$_{(A,B,C)}$ & cat. 5)$_{(A,B,C)}$ & seren. & Sum \\ 
\hline 
  no redshift & 57                 &20              &43               & 37            &0             &49                &206 \cr 
  stars       & 4$_{(0,3,1)}$      &0$_{(0,0,0)}$   &0$_{(0,0,0)}$    & 7$_{(1,2,4)}$ &0$_{(0,0,0)}$ &3$_{(2,0,1)}$&14 \cr 
  (0..1)      & 19$_{(12,1,6)}$    &42$_{(35,6,1)}$ &23$_{(18,3,2)}$  & 2$_{(2,0,1)}$ &1$_{(1,0,0)}$ &121$_{(83,19,18)}$&208 \cr 
  [1..2)      & 193$_{(113,51,29)}$&83$_{(49,24,10)}$&115$_{(76,25,13)}$& 4$_{(2,2,0)}$ &2$_{(2,0,0)}$ &34$_{(10,15,9)}$ &431 \cr
  [2..3)      & 0$_{(0,0,0)}$      & 1$_{(1,0,0)}$  & 0$_{(0,0,0)}$   & 0$_{(0,0,0)}$ &0$_{(0,0,0)}$ & 0$_{(0,0,0)}$    &1   \cr 
  [3..4)      & 0$_{(0,0,0)}$      & 0$_{(0,0,0)}$  & 0$_{(0,0,0)}$   &26$_{(14,7,4)}$&0$_{(0,0,0)}$ & 0$_{(0,0,0)}$    &26  \cr 
  [4..5)      & 0$_{(0,0,0)}$      & 0$_{(0,0,0)}$  & 0$_{(0,0,0)}$   &23$_{(6,8,9)}$ &0$_{(0,0,0)}$ & 2$_{(0,1,1)}$    &25  \cr 
  [5..6)      & 0$_{(0,0,0)}$      & 0$_{(0,0,0)}$  & 0$_{(0,0,0)}$   &14$_{(7,3,5)}$ &0$_{(0,0,0)}$ & 3$_{(0,1,2)}$    &17  \cr 
  [6..7)      & 0$_{(0,0,0)}$      & 0$_{(0,0,0)}$  & 0$_{(0,0,0)}$   & 2$_{(0,1,1)}$ &0$_{(0,0,0)}$ & 0$_{(0,0,0)}$    &2   \cr 
\hline
   Sum        & 273                &146             & 181              &115           &3             &212           &$\bf{930}$\cr   
\hline                                                                                             
\hline
\label{tab:matrix} 
\end{tabular} 
\end{table*}

\subsection{Redshift distribution and Large Scale Structure} 
The top and bottom panels of Figure~\ref{fig:zdistr} show the redshift 
distribution of the galaxies at redshift less than 2 and greater than 2, respectively 
(solid line QF ``A'' and ''B'', dotted line QF ``C''). In the following sections we 
discuss the redshift distribution separating the low ($z<2$) and
high ($z>2$) redshift intervals.

\subsubsection{The sample at $z < 2$}
The redshift distribution is consistent with the criteria 
for the target selection (color and photometric redshift selected),
with the majority of the sources having redshifts in the interval 1$<$z$<$2
(see also Table~\ref{tab:matrix}, last column).
In the RUN1+RUN2, out of 181 galaxies selected via photometric redshift, 138
have a spectroscopic redshift identification and 136 with zspec$>$0.8 (115 at zspec$>$1).

Table~\ref{tab:z_properties} shows the fraction of determined redshifts as 
a function of the spectral features identified, i.e. emission lines, absorption lines, 
emission \& absorption lines. The median of the redshift distribution of each class is close to 1,
with a more populated tail in the redshift interval 1$<z<$2 (see top panel of Figure~\ref{fig:zdistr}). 
\begin{figure}
 \centering 
 \includegraphics[width=8.5cm,height=9cm]{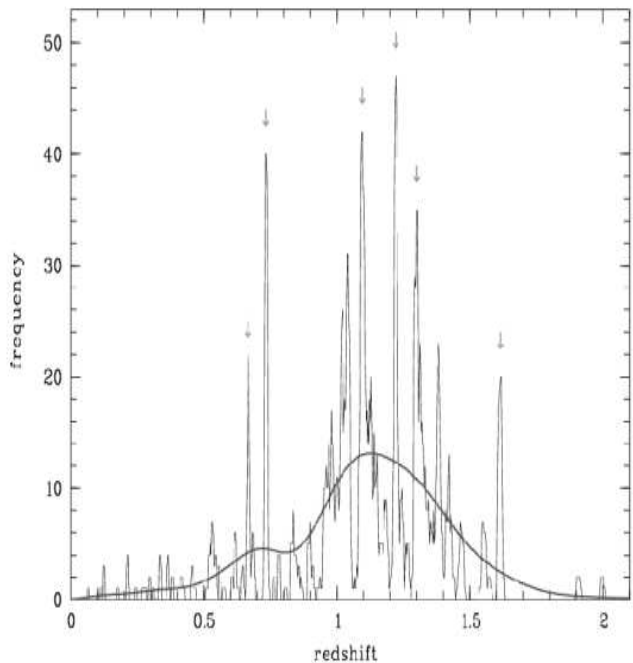}
 \caption{Redshift distribution of the spectroscopic sample at
 $z<2$. The signal has been smoothed with a Gaussian filter with
 $\sigma_{S}=300~km/s$ (the typical error in the redshift
 determination). The histogram has been obtained counting the number
 of sources in a window of $2000~km/s$ moved from redshift 0 to 2 with
 a step of $100~km/s$. The smoothed line is the ``background'' field
 distribution, obtained smoothing the observed distribution with a
 Gaussian filter with $\sigma_{S}=15000~km/s$. The peaks detected at
 $SNR>$5 are marked with an arrow (see text). Other two structures
 have been detected with a $SNR\sim$4.5 at redshift 1.040 and 1.382.}
\label{fig:LSS}
\end{figure}

Obviously, in the presence of emission lines, it is easier to determine a redshift.
As reported in Table~\ref{tab:z_properties}, 537 galaxies (including ``em.''
and ``comp'') show the  [O\,{\sc ii}]3727 emission line and assuming as an extreme case that
all the inconclusive redshift determinations (category ``X'') belong to the class ``abs.'',
the number of ``em.'' sources is still dominant, comprising 63$\%$ of the entire target list.
This is a likely reason why the majority of galaxies identified in the present
work belong to the ``em.'' class.
Alternatively, [O\,{\sc ii}]3727 is a classic star forming indicator and the redshift 
interval $1<z<2$ corresponds to the peak of the mean star formation intensity of the universe.

There are 102 galaxies identified with absorption lines only (``abs.'' class, mainly Ca H and K, 
MgII 2798-2802) in the range of redshift between 0.3-2.0. 
28 sources out of 102 with only absorption features detected have been 
serendipitously-observed, the redshift distribution of this sample peaks at z=0.68$\pm$0.2.
Six galaxies have been identified at redshift $\sim$2. These sources show 
the Mg\,{\sc ii} 2798,2802\AA~in absorption (in three cases the [Fe\,{\sc ii}] 2344,2383\AA~absorption
lines are also present), five of them 
(GDS~J033241.84-274657.1 QF=''B'', GDS~J033240.06-274755.4 QF=''A'', GDS~J033228.17-274648.4 QF=''C'', 
GDS~J033240.27-274949.7 QF=''C'' and GDS~J033233.84-274520.5 QF=''C'') 
have been discovered in the RUN2 and have blue colors ($i_{775}-z_{850}<0.6$).
Two examples of 2D spectra are shown in Figure~\ref{fig:GAL_Z2} and the composite 
one-dimensional spectrum is shown in the right panel of Figure~\ref{fig:stack1p61}.
For these sources the [O\,{\sc ii}]3727 emission, if present, is out of the spectral range, 
at 11180\AA. The source GDS~J033233.85-274600.2 is an elliptical galaxy at $z$=1.91 already 
discussed by \cite{cimatti04}, and has been observed in the RUN1.

\begin{table*} 
\centering \caption{Fractions of sources in the redshift interval 0$<$z$<$2 with different spectral features 
(RUN1 + RUN2 without stars). The fractions of the different categories observed in Sect.~2 are also shown.}
\begin{tabular}{lccc|cccccc|c}
\hline \hline
Spectral class &$(z_{median})_{-1\sigma}^{+1\sigma}$&$z_{min}$&$z_{max}$& cat.1)  & cat.2) & cat.3) & cat.4) &cat.5)   & cat.-1) (seren.)&Sum\\ 
\hline 
\cr
 emission      & (1.13)$_{-0.74}^{+1.33}$    & 0.067   & 1.621   & 117     &  92    &  123   &  3     &   2     &  104  & 441\cr 
\cr
 absorption    & (1.00)$_{-0.67}^{+1.22}$    & 0.337   & 1.998   & 51      &  15    &  5     &  2     &   1     &  28   & 102\cr 
\cr
 em. \& abs.   & (1.02)$_{-0.67}^{+1.29}$    & 0.382   & 1.380   & 44      &  18    &  10    &  2     &   0     &  22   & 96\cr 
\\
\hline
  Sum          &          &         &         &  212    &  125   & 138    &  7     &   3     &  154  & $\bf{639}$ \cr 
\hline 
\label{tab:z_properties} 
\end{tabular} 
\end{table*}

441 sources belong to the ``em.'' class (they are dominated by emission lines, mainly [O\,{\sc ii}]3727),
many of them entering the so-called ``spectroscopic desert'' up to z=1.621.
It is interesting to note that 133 galaxies out of 138 with redshift and photo-z selected show the  
[O\,{\sc ii}]3727 emission line.

96 sources have been classified as intermediate between  ``em.'' and ``abs.'' classes,
where both emission and absorption lines with an evident 4000\AA~break are present.

\subsubsection{Large Scale Structure}

The presence in the CDF-S of large scale structure (LSS) at $z<2$ is indicated by the 
peaks in the redshift distribution (see Figure~\ref{fig:LSS}). 
To assess the significance of these structures we follow a procedure similar to that
adopted by \cite{gilli03}, who observed features in their X-ray source redshift 
distribution.

We have distributed the sources (the ``signal distribution'') in the velocity domain 
($V~=~c~ln(1+z)$, so that $dV=\frac{c~dz}{1+z}$) and smoothed with a Gaussian filter with $\sigma_{S}=300~km/s$ 
(the typical error in the redshift determination).
We have then smoothed the observed distribution with a Gaussian filter with $\sigma_{S}=15000~km/s$
and considered this as the background distribution.

We have searched for possible redshift peaks in the signal distribution, computing a 
signal-to-noise ratio defined as $SNR$ = ($\frac{S-B}{B^{0.5}}$), where $S$ is the number 
of sources in a velocity interval of fixed width $\Delta V=2000~km/s$ and $B$ is the number 
of background sources in the same interval.
Adopting the threshold $SNR>$5 we have found 6 peaks in the signal distribution 
(indicated with an arrows in the Figure~\ref{fig:LSS}).

In order to estimate the expected fraction of possibly ``spurious'' peaks arising
from the background fluctuations, we have simulated 10$^{5}$ samples of the same size
of the observed distribution and randomly extracted from the smoothed background distribution 
and applied our peak detection method to each simulated sample.
The result is that, with the adopted threshold, the average number of spurious peaks
due to background fluctuations is 0.06. Of the simulated samples, 5.7$\%$ show one spurious 
peak, 0.1$\%$ show two spurious peaks, and only one simulation (out of $10^{5}$) has three
spurious peaks. None of the simulated samples have four or more spurious peaks.

\begin{table} 
\centering \caption{Peaks detected in the FORS2 source redshift distribution, sorted by increasing
redshift. The signal and background distributions are smoothed with $\sigma_{S}=300km/s$
and $\sigma_{B}=15000km/s$, respectively. Together with the central redshift of each peak,
the number of sources N in each peak and the probability (determined on 10$^{5}$ simulations) to 
detect spurious peaks arising from the background distribution with a $SNR$ equal or greater than 
the $SNR$ value measured in the signal distribution.} 
\begin{tabular}{ccccc} 
\hline \hline 
   z    &  N & SNR & Prob. \\ 
\hline 
  0.666 &  22 & 8.6   & 4.5$\times$10$^{-4}$  \cr 
  0.734 &  40 & 16.6  & $<$1$\times$10$^{-5}$ \cr 
  1.096 &  42 & 8.0   & 1.9$\times$10$^{-3}$  \cr 
  1.221 &  47 & 9.7   & 2.2$\times$10$^{-4}$  \cr 
  1.300 &  35 & 7.4   & 4.2$\times$10$^{-3}$  \cr 
  1.614 &  20 & 11.1  & 7.0$\times$10$^{-5}$  \cr 
\hline
\label{tab:LSS}
\end{tabular}
\end{table}

The six source peaks detected by our procedure are listed in Table~\ref{tab:LSS}, where for each
peak we give the average redshift, the number of objects (N) in the peak and the 
probability (derived from the 10$^{5}$ simulations) of observing a spurious peak with the SNR equal or 
greater than the measured SNR of the peak detected in the signal distribution.

\begin{figure}
 \centering
 \includegraphics[width=9cm,height=9cm]{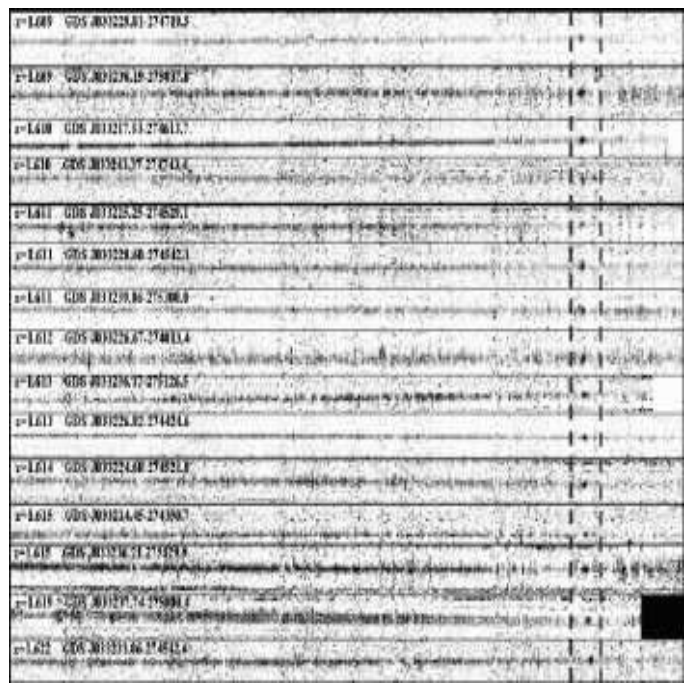}
 \caption{Two dimensional spectra of 15 galaxies at z$\sim$1.61 discovered
in the RUN2. The [O\,{\sc ii}]3727 emission line is marked with vertical dashed lines
around the 9727.5\AA~position.}
\label{fig:groupz1.61}
\end{figure}
\begin{figure*}
 \centering
 \rotatebox{0}{
 \includegraphics[width=9cm,height=9cm]{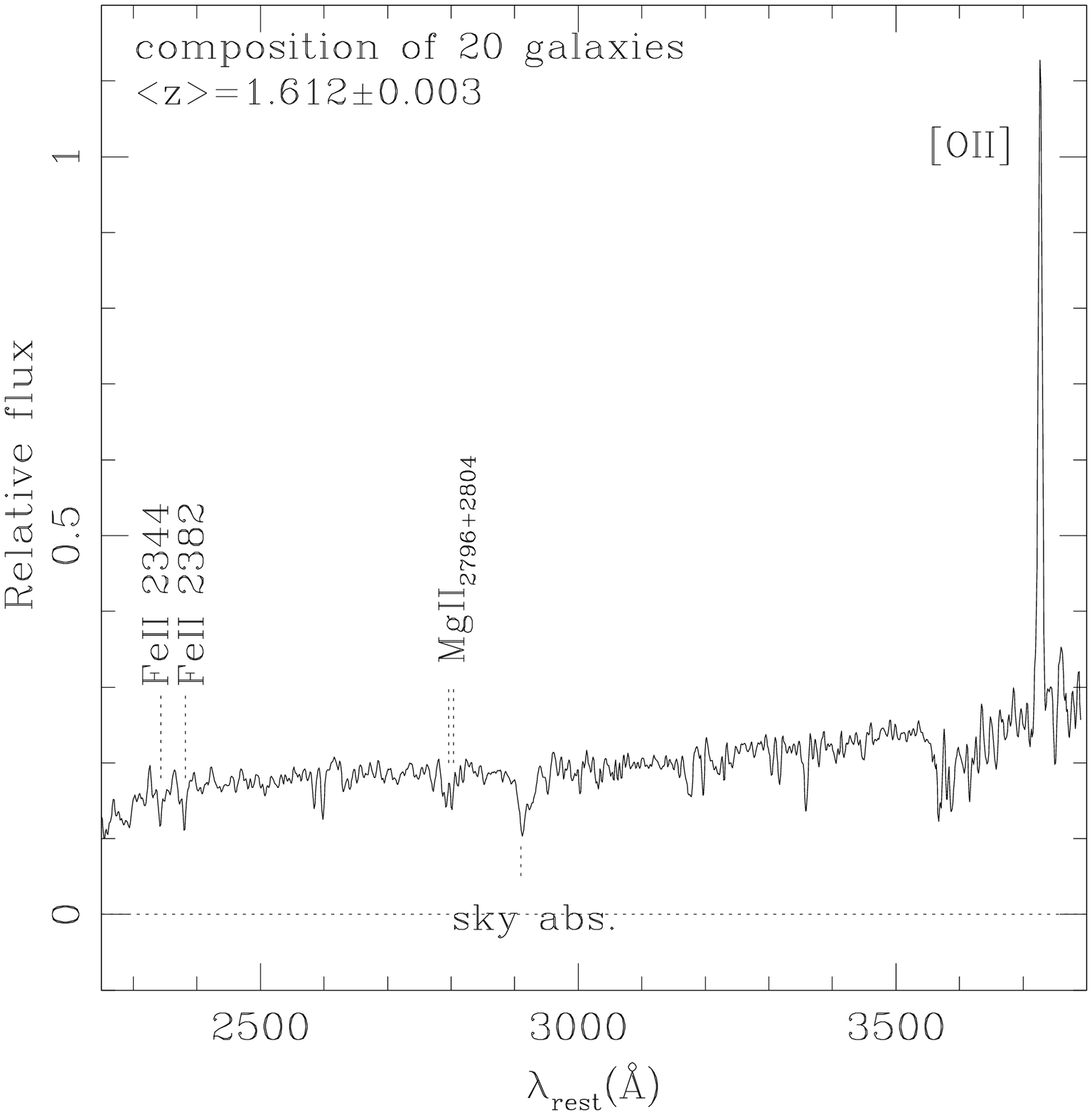}
 \includegraphics[width=9cm,height=9cm]{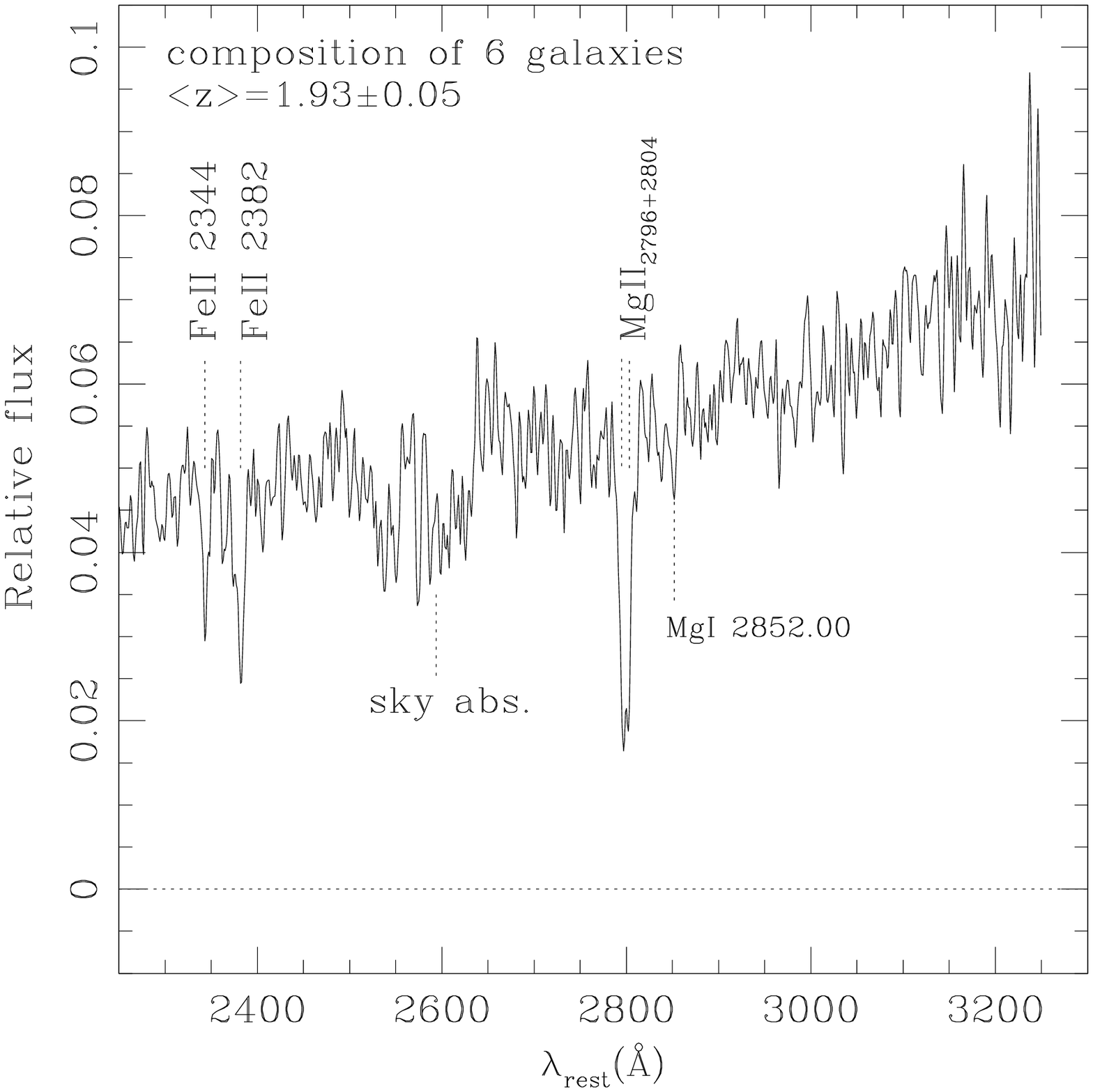}
 }
 \caption{Left panel: rest frame composite spectrum of 20 galaxies at redshift $\sim$1.61. The [O\,{\sc ii}]3727 
is clearly evident and the Mg\,{\sc ii} 2798,2802\AA~and  [Fe\,{\sc ii}] 2344,2383\AA~are also present.
Right panel: rest frame composite spectrum of 6 galaxies at redshift $\sim$1.93.}
\label{fig:stack1p61}
\end{figure*}

The peaks at $z \sim 0.734$ and $z \sim 0.666$ are
already known (\cite{cimatti02}, \cite{gilli03}, \cite{fevre04}).
The other four indications of large scale structures in the CDF-S have been identified at 
redshift 1.096, 1.221, 1.300 (also described by \cite{adami05}) and 1.614.
We note that other two peaks have been detected with a $SNR\sim$4.5 at redshift 1.040 and 1.382.

In the current spectroscopic catalog (RUN1 + RUN2) 20 galaxies at $z \sim 1.61$ have
been discovered (Figure~\ref{fig:groupz1.61} shows an example of the
z$\sim$1.61 galaxies discovered in the RUN2). 
The number of sources increase if we consider other surveys: 
\begin{enumerate} 
\item{the observations of \cite{gilli03} who found a peak in the redshift distribution of X-ray sources 
at z=1.618 (5 galaxies);} 
\item{the three galaxies at z$\sim$1.61 (\cite{cimatti02}, \cite{cimatti04}) which are passively
evolving early type galaxies}
\item{at least 5 more galaxies in the third FORS2 run (from our preliminary reduction);}
\end{enumerate} 
\begin{figure} 
 \centering 
 \includegraphics[width=9cm,height=11cm]{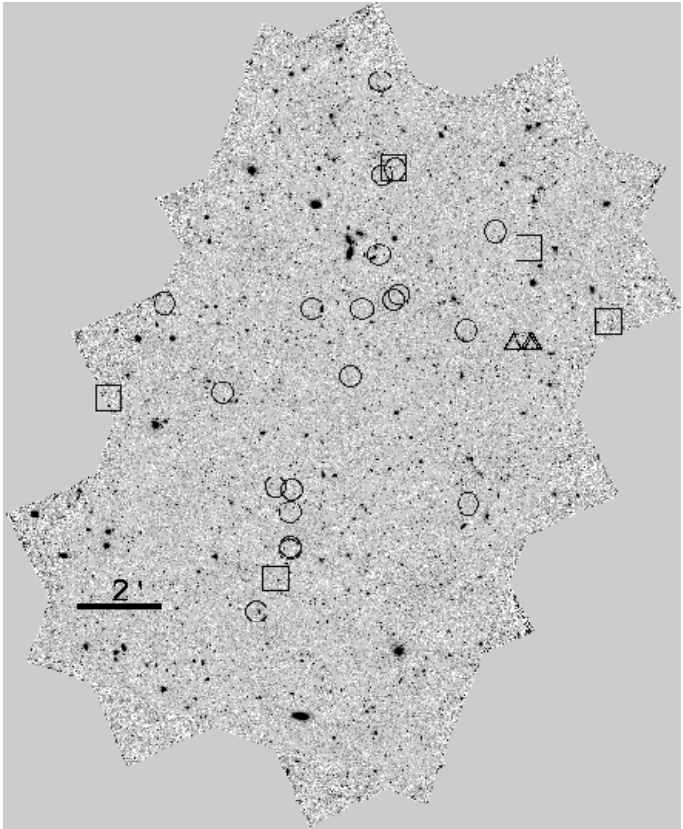}
 \caption{The spatial distribution of the galaxies at $z \sim 1.61$ in
the CDF-S. The background image is the ACS $z_{850}$ band GOODS field,
north is up and east on the left.
The horizontal bar indicates a 2 arcmin size (corresponding to 1020 kpc at z=1.61).
Twenty sources identified by FORS2 are marked with circles.
The three triangles mark the position of the K20 sources.
The squares show the positions of the $z \sim 1.61$ X-ray
sources (see text)}.
\label{fig:space_distr}
\end{figure}
 
Fig.~\ref{fig:space_distr} shows the spatial distribution of the galaxies at $z ~\approx~ 1.61$ using
both the present work and data from the literature.
The current sample contains 28 galaxies apparently distributed in a non-uniform way, the majority of them
have been detected in the upper part of the field and 3 pairs have an angular separation below 
4 arcseconds ($\sim$ 30 kpc at $z \sim 1.61$).

At redshift 1.61, the ACS $B_{435}$ band is sampling the 1667\AA~rest-frame UV radiation. As reviewed by
\cite{kenni98}, one can estimate the SFR from the rest-frame UV luminosity density $L_{\nu}$ in the
range 1500-2500~\AA~using the following relation: SFR(M$_{\odot}$yr$^{-1}$) = 1.4 $\times$ 10$^{-28}$
$L_{\nu}$ (ergs s$^{-1}$ Hz$^{-1}$) for a Salpeter IMF, covering the range 0.1 to 100M$_{\odot}$.
This relation applies only to galaxies with continuous star formation over time scales of 10$^{8}$ years
or longer.

We have estimated the rest-frame luminosity density $L_{\nu}$ (in ergs s$^{-1}$ Hz$^{-1}$) of the 20
galaxies at z=1.61 identified in the current FORS2 spectroscopic campaign, using the
apparent $B_{435}$ AB magnitude (the SExtractor ``mag$\_$auto'', \cite{ber96}) and the luminosity distance.
The final luminosity is $L_{\nu,o}$ = $L_{\nu}$ $\times$ 10$^{0.4~Av}$, where $Av$ represents the
amount of dust extinction. Adopting no extinction ($Av$=0), we obtain a lower limit for the mean
star formation rate of $<$SFR$>$4$\pm$2 M$_{\odot}$yr$^{-1}$. Assuming $Av$=1 or $Av$=2 the $<$SFR$>$ 
increase from 10$\pm$5 M$_{\odot}$yr$^{-1}$ to 24$\pm$14 M$_{\odot}$yr$^{-1}$, respectively.

The rest frame composite spectrum of twenty galaxies at z=1.61 is shown in the left panel of 
Figure~\ref{fig:stack1p61}.
The [O\,{\sc ii}]3727 line and the Mg\,{\sc ii} 2798,2802\AA~and [Fe\,{\sc ii}] 2344,2383\AA~are
clearly evident.

\subsection{The Lyman break galaxies}
116 sources in the FORS2 RUN1 and RUN2 belong to the class 4), i.e., objects selected to be at high
redshift by Lyman break color criteria. It is important to divide the first and second run in 
order to characterize the success rate. 
As already discussed in the previous paper (\cite{vanz05}), in the first FORS2 run
14 candidate dropouts were observed, and only one was confirmed at z=5.83. Another five were found to be
stars and the remaining sources had inconclusive spectra. The photometric
selection of the dropouts galaxies in the first FORS2 run was based on an incomplete photometric
dataset (first three epochs photometry).

In the following, we consider only the results from RUN2, for which dropout
candidates were selected from the full (five epochs) ACS photometry.

94 Lyman break galaxy candidates selected by the $B_{435}$,$V_{606}$ and $i_{775}$-dropout criteria were
observed in RUN2. The redshift distribution measured for 65 of these galaxies is shown in the
lower panel of Figure~\ref{fig:zdistr}.
The 75$\%$, 70$\%$ and 70$\%$ of the observed $B_{435}$, $V_{606}$ and $i_{775}$-dropout color
selected candidates have a redshift estimation. The sources with inconclusive redshift determination
are in general too faint or without evident spectral features.

100$\%$ of the $B_{435}$-dropouts with a measured redshift have been confirmed to be at redshift between
3.4 and 4.6, 90$\%$ of the $V_{606}$-dropouts with a measured redshift are in the range 4.4 and 5.6,
and 93$\%$ of the $i_{775}$-dropouts with a measured redshift are at redshift greater than 5.2
(one source is a probable star).

\begin{table*} 
\centering \caption{Fraction of confirmed dropout candidates in the second FORS2 run (RUN2), ``Nobs.'' indicates 
the number of sources observed. Four serendipitously-observed high redshift sources are also reported.}
\begin{tabular}{lccc|ccc} 
\hline \hline 
 classes (Nobs.) & confirmed high-z (*)& confirmed low-z  & no redshift & (*) ``em.''$_{(A,B,C)}$ &(*)  ``abs.''$_{(A,B,C)}$ &(*)  ``em.''+''abs.''$_{(A,B,C)}$ \\ 
\hline 
  $B_{435}$-drop (44)   & 33 (3.418$<$z$<$4.597)& 0            &  11  &12$_{(8,3,1)}$ &20$_{(9,5,6)}$ &1$_{(1,0,0)}$\cr
  $V_{606}$-drop (30)   & 19 (4.400$<$z$<$5.554)& 2 (z$<$1.4)  &  9   &13$_{(5,5,3)}$ & 6$_{(0,2,4)}$ &0$_{(0,0,0)}$\cr
  $i_{775}$-drop (20)   & 13 (5.250$<$z$<$6.200)& 1 (star)     &  6   & 8$_{(3,4,1)}$ & 5$_{(0,0,5)}$ &0$_{(0,0,0)}$\cr
  Serend.               & 3 (4.838$<$z$<$5.541)  &  -   &-& 3$_{(0,0,3)}$ & 0$_{(0,0,0)}$ &0$_{(0,0,0)}$\cr
\hline
\label{tab:high-z} 
\end{tabular} 
\end{table*}  

Table~\ref{tab:high-z} (and Table~\ref{tab:matrix}) summarize the success rate as a function of 
redshift, quality flag, class and selection criteria. 
Columns 5, 6 and 7 of Table~\ref{tab:high-z} show the fraction of the confirmed high redshift galaxies
and the ``class'' flag that is related to the features detected in the redshift determination.
Beyond redshift 5, if no spectral lines are present,
the main features indicating the high redshift nature of the source are: the break in the continuum 
due to galactic and intergalactic absorption blueward 1215.8\AA, and the flatness of the continuum 
redward the 1215.8\AA. 

\begin{figure} 
 \centering
 \includegraphics[width=9cm,height=13cm]{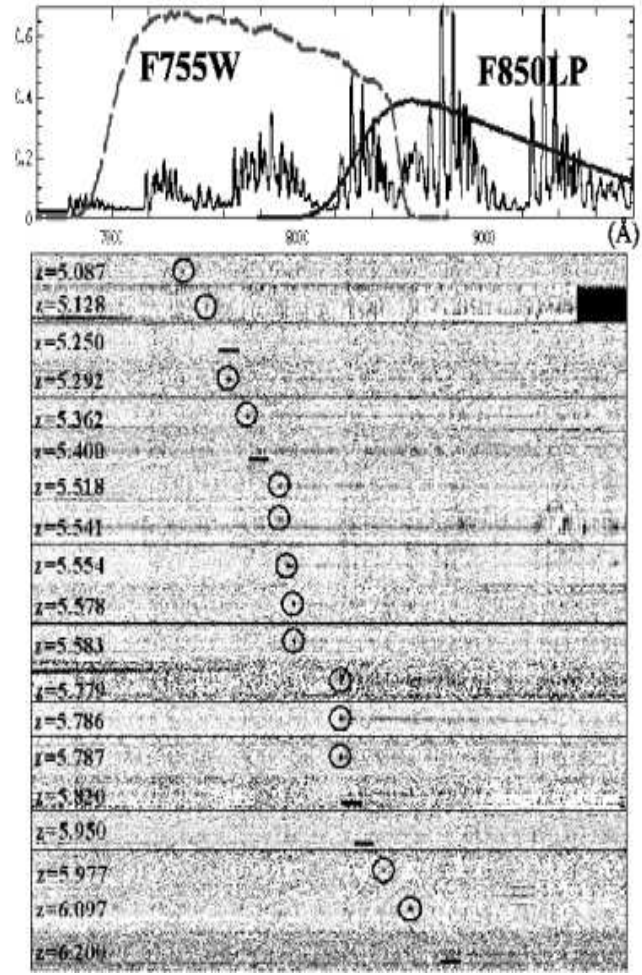}
 \caption{Two dimensional spectra of 18 galaxies at redshift greater than 5 discovered in the FORS2 campaign. The
$Ly{\alpha}$ line is marked with a circle. For the other cases the segment outline the possible position of
the break. The 1-D spectra of the GOODS/FORS2 Data Release version 2.0 are available at the following
URL: {\it http://www.eso.org/science/goods/}.}
\label{fig:panoramic_high_z} 
\end{figure} 

\begin{figure}
 \centering
 \includegraphics[width=9cm,height=9cm]{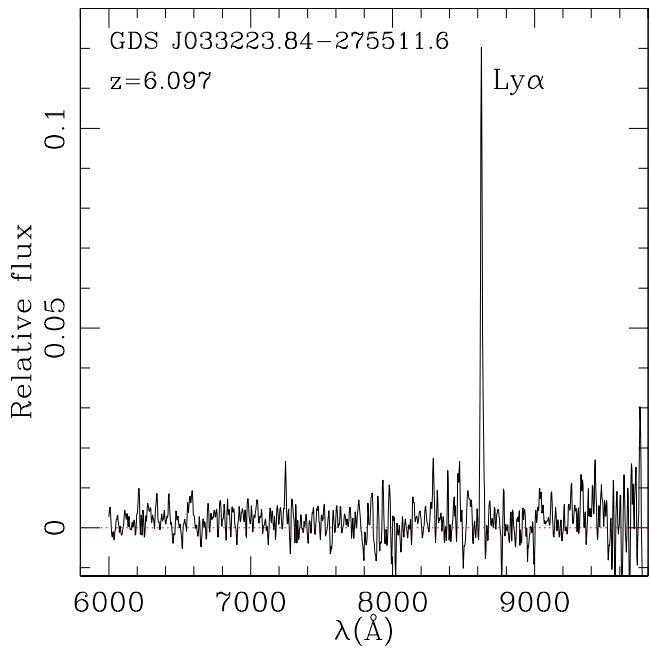}
 \caption{One-dimensional spectrum of the $i_{775}$-dropout galaxy at redshift 6.097 (QF=''B'').}
\label{fig:z6p097}
\end{figure}

Figure~\ref{fig:panoramic_high_z} shows the two-Dimensional collection of the 18 galaxies at 
redshift greater than 5 discovered in the RUN2 and Figure~\ref{fig:z6p097} shows the one-dimensional
spectrum of the galaxy GDS~J033223.84-275511.6 at z=6.097.

In the top of the figure the spectrum of the sky (not flux calibrated)
is shown together with the response curves of the ACS filters $i_{755}$ and the $z_{850}$.
In some cases the Ly$\alpha$ is in emission (marked with a circle)
and the break of the continuum is evident.
Five sources show only the continuum break
(a solid segment marks the possible position of the break).
The mean value of the observed $i_{755} - z_{850}$ for this sample increases with
increasing redshift.
The presence of the Lyman emission line, however,
can affect significantly the resulting color of the galaxy, introducing a scatter in 
the blue or in the red directions. For example in the case of the source 
GDS~J033218.92-275302.7, the strong Ly$\alpha$ line at z=5.554 produces an 
$i_{755} - z_{850}$ = 0.625.
Similarly, in the case of the source GDS~J033223.84-275511.6, the intense Ly$\alpha$ line
falls in the $z_{850}$ band, producing an $i_{755} - z_{850}$ $>$ 4.

\begin{figure*} 
 \centering 
 \includegraphics[width=15cm,height=10cm]{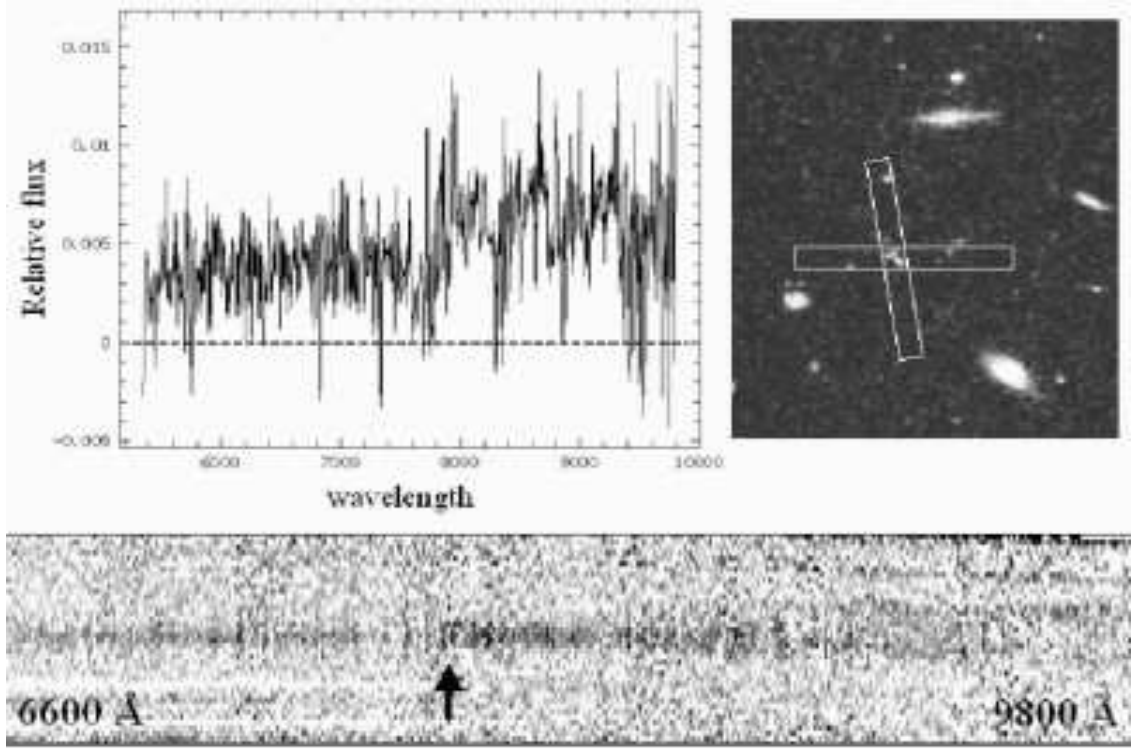}
 \caption{One and two dimensional spectra of the $i_{775}$-dropout candidate GDS~J033217.96-274817.0
 (observed in two masks).  
 In the upper right side of the figure, the ACS color image and the orientation of the slits are shown.
 The intruder GDS~J033217.95-274817.5 is a blue source with $i_{775}-z_{850}$=-0.25, while the 
 $i_{775}$-dropout candidate is clearly red, $i-z$=1.18. 
 In the one-dimensional spectrum, the break at 
 $\sim$ 7800\AA~(indicated with the arrow in the two dimensional spectrum) and the flatness shape 
 redward the break are shown. 
 Collapsing $\sim$ 100 columns below and beyond the 7800\AA~break, the two resulting profiles 
 are shifted of $\sim$ 0.4 arcsecond, consistently with the separation of the two sources measured in the 
 ACS image. Assuming the 7800\AA~break as due to the high redshift nature of the redder source, 
 the redshift turns out to be $\sim$ 5.4 (quality flag ``C'').}
\label{fig:idrop_blue} 
\end{figure*}

The source GDS~J033217.96-274817.0 is an $i_{775}$-dropout candidate. 
The FORS2 spectrum is the superposition of two sources. 
In Figure~\ref{fig:idrop_blue} the one and two dimensional spectra and the 
color ACS image of the sources are shown. One source (GDS J033217.95-274817.5)
is clearly blue ($i_{775}-z_{850}$=-0.25) with respect the $i_{775}$-dropout candidate ($i-z$=1.18). 
The one dimensional spectrum shows a break at $\sim$ 7800\AA~and the flatness shape redward the break. 
Collapsing $\sim$ 100 columns below and beyond the 7800\AA~break, the two resulting profiles are shifted 
of $\sim$ 0.4 arcsecond, consistently with the separation of the two sources measured in the ACS image.
Interpreting this break due to the high redshift nature of the $i_{775}$-dropout source, the redshift 
is $\sim$ 5.4 with QF=''C''. We note that the uncertainty of the position of the break
is increased by the presence of the sky absorption A-band at $\sim$7600\AA.
\subsection{Galaxies showing a tilted [O\,{\sc ii}]3727 line}
\begin{figure}
 \centering
 \includegraphics[width=9cm,height=12cm]{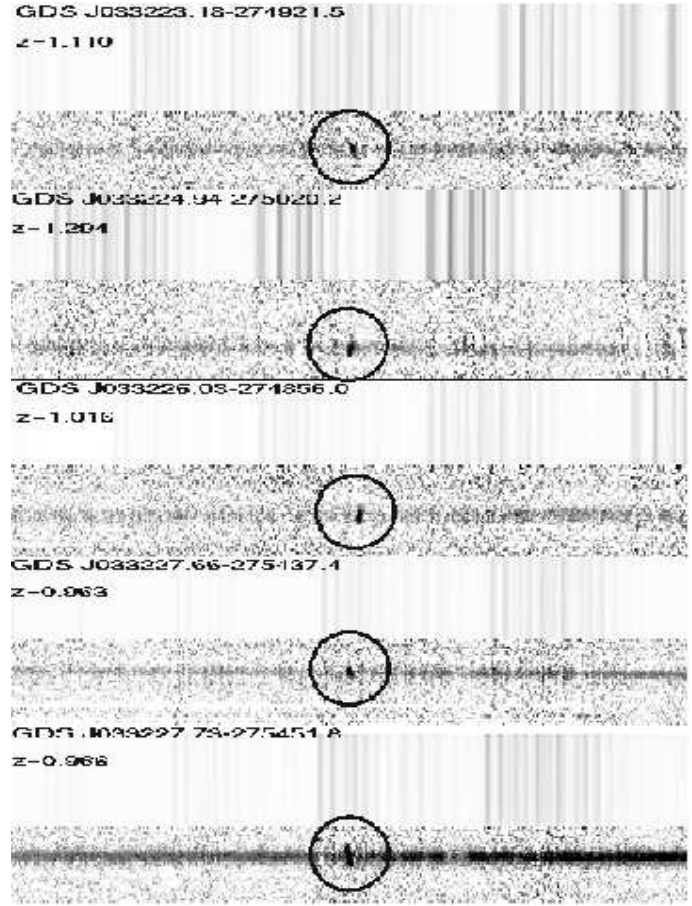}
 \caption{An example of tilted [O\,{\sc ii}]3727 emission line (marked with a circle) at 
redshift around 1. The two dimensional FORS2 spectra are shown (object and sky lines).} 
\label{fig:OIItilted1} 
\end{figure}

The current FORS2 spectroscopic catalog contains a sample of sources showing a spatially
resolved [O\,{\sc ii}]3727 line with a characteristic ``tilt'' indicative of a high
rotation velocity. Table~\ref{tab:OII_TILT} lists the 34 sources sorted by increasing redshift, 
the majority of them belong to the interval 1$<$z$<$1.5.
Figure~\ref{fig:OIItilted1}
show an example of the two dimensional spectra of the galaxies and the sky lines. The [O\,{\sc ii}]3727 line
is marked with a circle.

As discussed in the previous paper (\cite{vanz05}) the resolution of the FORS2 spectra
favor the detection of high velocity rotational systems. Moreover 
a not optimized orientation of the slit suggest that in general the ``true'' maximum velocities 
may be significantly higher than the measure value.

\begin{figure*}
 \centering
 \rotatebox{0}{
 \includegraphics[width=9cm,height=9cm]{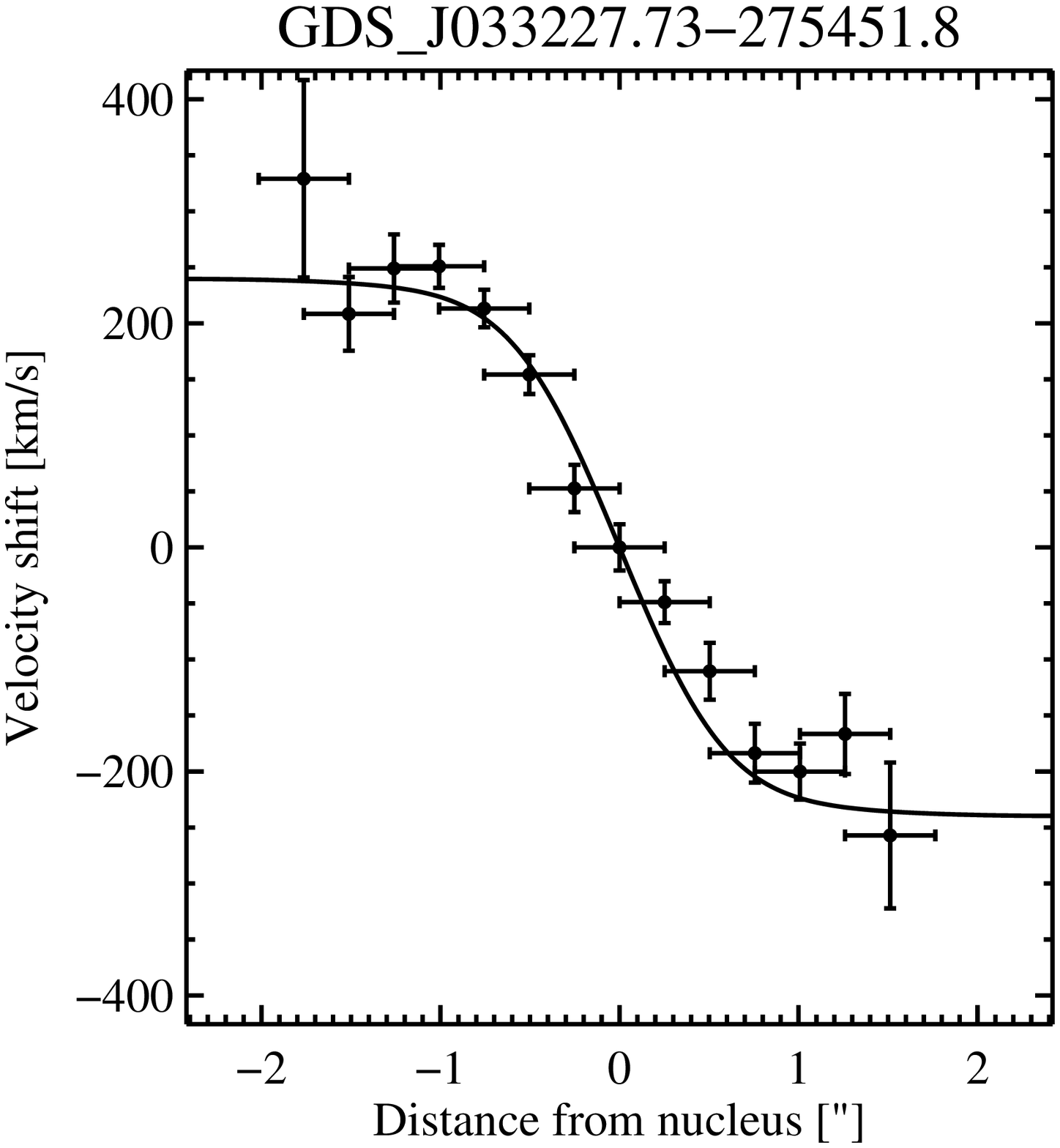}
 \includegraphics[width=9cm,height=9cm]{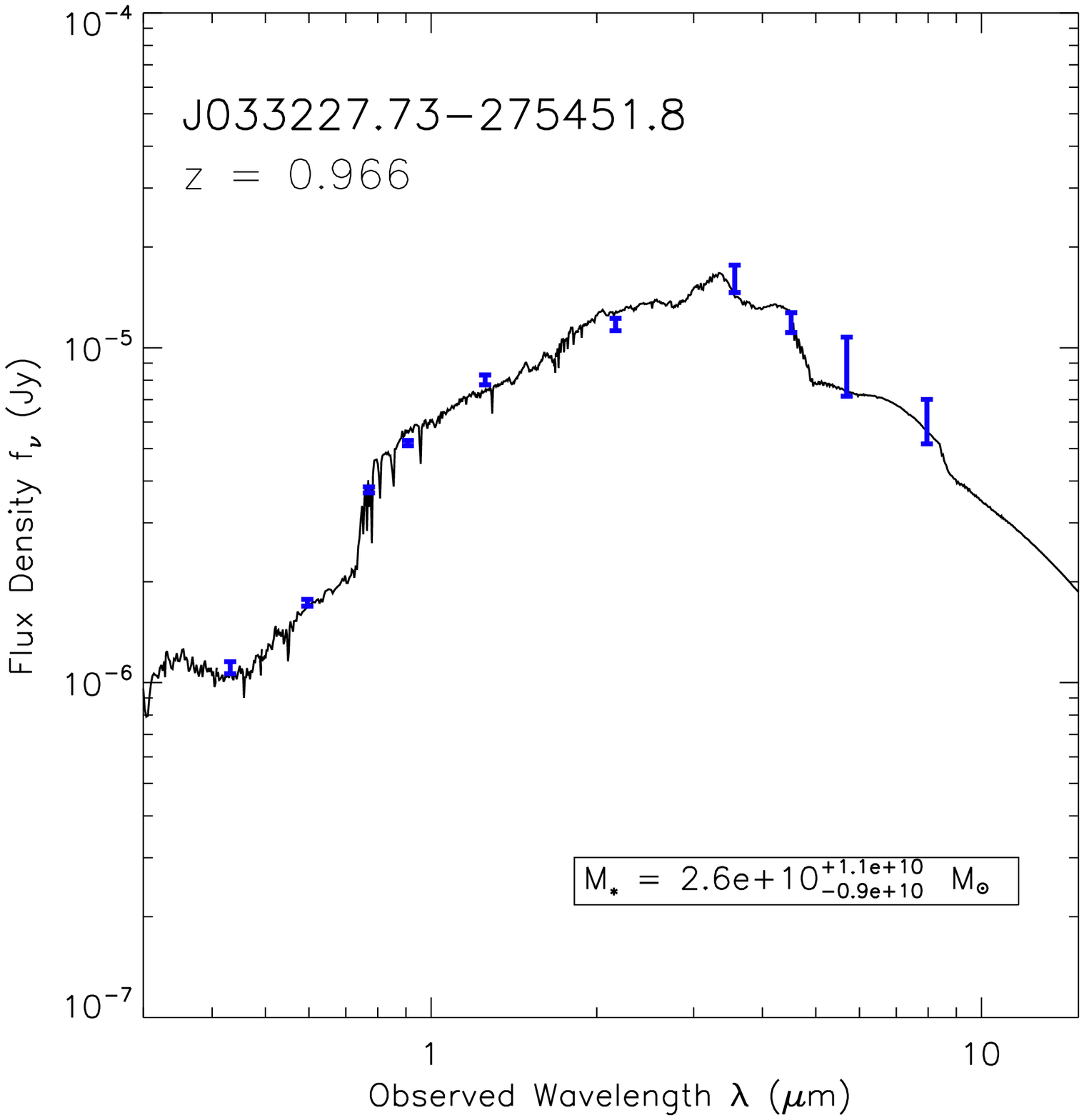}
 }
 \caption{An example of dynamical mass estimation from the analysis of the rotational curve, 
  compared with the photometric stellar mass estimation obtained through the SED fitting technique 
  for one of the sources with tilted  [O\,{\sc ii}]3727 emission line (z=0.966). The total
  halo mass is $1\pm0.4 \times10^{12}M\sun$ and the stellar mass is $2.6^{+1.1}_{-0.9}\times10^{10}M\sun$.}
\label{fig:Vrot}
\end{figure*}

 As an example, we have analyzed the velocity field of object 
 J033227.73-275451.8 and estimate the stellar mass from the multi-wavelength dataset. 

 We first traced the centroid of the 
 [O\,{\sc ii}]$\lambda\lambda$3726,3729 emission line doublet along the spatial 
 position. Since the resolution of our spectrum is too low to resolve the 
 doublet, we fixed the ratio between the two components to 1 and we 
 checked that the results were fairly insensitive to this assumption.
 We then compared this measured rotation curve with a set of synthetic 
 rotation curves for which the velocity rises linearly up to one disk 
 scale length and is flat at larger radii. This step takes into account 
 the inclination of the disk with respect to the line of sight ($i=53\pm3 
 \deg$), the disk scale length ($r_d=0.405 \pm 0.045 \arcsec$) 
 (derived from the morphological analysis, Rettura et al., in preparation), 
 the slit misalignment with respect to the galaxy major axis ($22\pm2 \deg$), 
 the width of the slit (1") and the seeing ($0.7\arcsec$) (this 
 method is similar to that of \cite{boehm04}). The best fit rotation 
 curve gives a rotation velocity of $355\pm50$ {\it km$s^{-1}$}. Using 
 prescription from \cite{bosch02}, this implies a total halo mass of 
 $1\pm0.4 \times10^{12}M\sun$.

Dedicated spectroscopic observations specifically designed 
for the dynamical mass estimation (higher spectral resolution, optimized slit orientation, 
etc.), should be performed/preferred in order to decrease the uncertainties.

We have used the full optical (HST/ACS B,V,i,z), near infrared (VLT/ISAAC J,Ks) to mid-infrared 
(Spitzer/IRAC $3.6 \mu, 4.5 \mu, 5.8 \mu, 8.0 \mu$) data to study the Spectral Energy 
Distribution (SED) of the same galaxy.
We have adopted $1.5~arcsec$ radius aperture-corrected to $3.5~arcsec$ radius photometry
to account for different instrumental PSFs. We have compared the observed SED with a set of
template computed with P\'EGASE.2 models (\cite{RV97}) via $\chi^{2}$ minimization technique
(the Salpeter IMF has been assumed). A more detailed description of the multi-wavelength cataloging and the
fitting SED technique used here will be presented in Rettura et al. (in preparation) on a
larger sample. We have calculated the errors for the mass estimate by sampling the full 
probability distribution in the parameters-space. Results of the SED fit are shown in
Figure~\ref{fig:Vrot} right panel. We find a best-fit stellar mass of 
$2.6\times10^{10}~M_{\odot}$ with a 1$\sigma$ confidence interval between 
1.7-3.7$\times10^{10}~M_{\odot}$. 
The comparison between this estimation and the dynamical halo mass produces a 
stellar mass over halo mass ratio of $f*=0.026$. This result is consistent with the estimations
performed by \cite{conselice05} on a large sample of disk galaxies at $z\leq$1.1,
where they find a wide range of $f*$ values (0.004 $\leq f* \leq $ 2).

\begin{table} 
\centering \caption{Sample of 34 galaxies (RUN1 + RUN2) with tilted [O\,{\sc ii}]3727.}
\begin{tabular}{lcccc}
\hline \hline
 GDS ID & zspec& class & quality \\
\hline
    GDS J033254.87-275456.0     & 0.125 &em.    &  C \cr
    GDS J033237.54-274838.9     & 0.665 &em.    &  A        \cr
    GDS J033215.88-274723.1     & 0.896 &em.    &  A \cr
    GDS J033227.66-275437.4     & 0.963 &em.    &  A        \cr
    GDS J033227.73-275451.8     & 0.966 &comp.  &  A        \cr
    GDS J033249.73-275517.4     & 0.981 &em.    &  A        \cr
    GDS J033234.56-275543.6$\dag$& 0.983 &em.  &  A        \cr
    GDS J033222.44-275606.1$\dag$& 1.090 &em.  &  C        \cr
    GDS J033226.03-274856.0     & 1.016 &em.    &  A        \cr
    GDS J033230.50-275312.3     & 1.017 &comp.  &  A        \cr
    GDS J033225.28-275524.2     & 1.017 &comp.  &  A        \cr
    GDS J033235.72-275615.4     & 1.033 &em.    &  A        \cr
    GDS J033233.71-274210.2     & 1.043 &em.    &  B \cr
    GDS J033234.42-275405.7     & 1.088 &comp.  &  A        \cr
    GDS J033225.86-275019.7     & 1.095 &em.    & A \cr
    GDS J033246.71-274556.0     & 1.095 &em.    &  B        \cr
    GDS J033247.42-274711.1     & 1.098 &em.    &  A        \cr
    GDS J033215.23-274437.8     & 1.109 &em.    &  B        \cr
    GDS J033223.18-274921.5     & 1.110 &em.    &  B        \cr
    GDS J033216.28-274447.6     & 1.183 &em.    &  C        \cr
    GDS J033216.26-274703.3     & 1.219 &em.    &  A        \cr
    GDS J033238.01-275408.2$\dag$&      1.243 &em.    &  B        \cr
    GDS J033224.94-275020.2     & 1.294 &em.    &  B       \cr
    GDS J033205.67-274253.5     & 1.296 &em.    &  A       \cr
    GDS J033232.42-274150.1$\dag$&      1.296 &em.    &  B        \cr
    GDS J033232.47-274151.5$\dag$&      1.296 &em.    &  B        \cr
    GDS J033213.21-274158.0     & 1.297 &em.    &  B        \cr
    GDS J033240.94-274427.5     & 1.298 &comp.  &  A        \cr
    GDS J033244.35-275506.4$\dag$&      1.305 &em.    &  A        \cr
    GDS J033230.71-274617.2     & 1.307 &em.    &  A \cr
    GDS J033234.82-274721.9$\dag$& 1.316&em.    &  B \cr
    GDS J033239.66-275406.3     & 1.343 &em.    &  A        \cr
    GDS J033240.08-275532.6     & 1.461 &em.    &  A        \cr
    GDS J033229.06-275542.8     & 1.469 &em.    &  A        \cr
\hline
\multicolumn{4}{l}
{$\dag$ possible tilted line.}\\
\label{tab:OII_TILT} 
\end{tabular} 
\end{table}

\section{Conclusions} 
As a part of the Great Observatories Origins Deep Survey, a 
large sample of galaxies in the Chandra Deep Field South has been 
spectroscopically targeted. 
After the RUN1 (\cite{vanz05}) and RUN2 (present work) a total of 930 objects
with $z_{850} \mincir 26.8$ have been observed with the FORS2 spectrograph at the 
ESO VLT providing 724 redshift determinations.  
From a variety of diagnostics the measurement of the redshifts appears 
to be precise (with a typical $\sigma_z \simeq 0.001$) and reliable.
The reduced spectra and the derived redshifts are released to the community  
($\it{http://www.eso.org/science/goods/}$). 
They constitute an essential contribution to reach the scientific goals 
of GOODS, providing the time coordinate needed  
to delineate the evolution of galaxy masses, morphologies, and star 
formation, calibrating the photometric redshifts that can be derived from the 
imaging data at 0.36-8$\mu$m and enabling detailed studies  
of the physical diagnostics for galaxies in the GOODS field. 
 
\begin{acknowledgements} 
 We are grateful to the ESO staff in Paranal and Garching who greatly helped 
 in the development of this programme. 
 The work of DS was carried out at the Jet Propulsion Laboratory, 
 California Institute of Technology, under a contract with NASA. 
 We thank the ASI grant I/R/088/02 (SC, MN, EV). 
\end{acknowledgements}

\end{document}